\documentclass[preprint,aps,nofootinbib]{revtex4}
\usepackage{}
\usepackage{graphicx}
\usepackage{amsmath}
\usepackage{amsfonts}
\usepackage{amssymb}
\usepackage{color}%
\usepackage{dcolumn}
\usepackage{bm}
\usepackage{float}
\providecommand{\U}[1]{\protect\rule{.1in}{.1in}}

\newcommand{\f}{\begin{equation}}
\newcommand{\ff}{\end{equation}}
\newcommand{\fa}{\begin{eqnarray}}
\newcommand{\ffa}{\end{eqnarray}}

\begin{document}

\title{ Holographic p-wave Superconductors in Quasi-topological Gravity}
\author{Xiao-Mei Kuang $^{1}$}
\email{xmeikuang@gmail.com}
\author{Wei-Jia Li $^{2}$}
\email{li831415@163.com}
\author{Yi Ling $^{1}$}
\email{yling@ncu.edu.cn} \affiliation{$^1$Center for Relativistic
Astrophysics and High Energy Physics, Department of Physics,
Nanchang University, 330031, China\\$^2$Department of Physics, Beijing Normal University, Beijing 100875, China}

\begin{abstract}
We construct a holographic p-wave superconductor model in the
framework of quasi-topological gravity in the probe limit. The
relation between the critical temperature and the coupling
parameters of higher curvature terms is investigated. The numerical
results for conductivity are presented as well. It turns out that
our data fits the Drude model very well in the low frequency limit,
and the values of DC conductivity as well as the relaxation time are
obtained numerically.

\end{abstract} \maketitle

\section {Introduction}
Recently the gauge/gravity
duality\cite{ADS,WittenADS,MaldacenaReviewADS} has been widely
applied to the study of condensed matter physics. In particular,
some critical phenomena in strongly coupled systems can be described
by the dynamics of geometry as well as matter fields in a
semi-classical region (For recent reviews we refer to
\cite{Hartnoll,McGreevy,Sachdev,horowitz1}). One remarkable example
is building holographic superconductor models based on the Abelian
Higgs mechanism, through which an asymptotical anti-de Sitter black
hole can break the $U(1)$ gauge symmetry spontaneously
\cite{Gubser}. In these models, when the effective mass of a scalar
field in the bulk is below the Breitenlohner-Freedman bound
\cite{Breitenlohner}, a hairy solution to the scalar field can be
obtained, characterized by its condensation around the horizon of
the black hole . According to AdS/CFT correspondence, the creation
of charged condensation strongly implies that a second order phase
transition could occur in the dual
CFT\cite{Hartnoll2,3H,Herzog,horowitz3}. Furthermore, this sort of
holographic superconductors contain more interesting
features\cite{bin1,bin2,bin3,bin4,hb,wjp2,Csb1,Csb2,Kanno,bin5,bin6,Gregory}.
For instance, the energy gap is much larger than the predictions of
the conventional BCS theory but quite similar to the high-$T_c$
superconductors as found in experiments\cite{Gomes}.

In addition to various holographic s-wave superconductor models with
scalar fields, one can also construct p-wave superconductor models
with vector fields, implemented by introducing a SU(2) Yang-Mills
gauge field\cite{Gubser2,Gubser3,cai,Ammon,Zeng,cai2}.
Correspondingly, the dual CFT has a global SU(2) symmetry and hence
a conserved current $J_{\mu}^{i}$. In the AdS black hole background,
it is found that the $U(1)_3$ gauge symmetry (a subgroup of $SU(2)$)
is possibly spontaneously broken such that the value of $\langle
J_x^1\rangle$ is not vanishing. This corresponds to a phase
transition between a non-superconducting state at high temperature
and a superconducting state below the critical temperature. However,
different from the s-wave model, here the order parameter is a
current and the conductivity is anisotropic since the condensation
of vector field breaks the rotational symmetry as well. As a result
the conductivity has two independent components. One perpendicular
to the direction of the condensation behaves like that in s-wave
superconductors, while the other parallel to it performs a much
different behavior. In particular, this component agrees well with
the Drude model in the low frequency limit.

Recently a gravity theory with nontrivial curvature-cubed terms in
five-dimensional spacetime, usually called quasi-topological
gravity, has been proposed in
\cite{Myers,oliva,ray,sinha1,sinha2}. This theory can be viewed as
a generalization of Gauss-Bonnet gravity. Besides the GB term in
the action, it also involves in higher-derivative corrections,
thereby corresponding to CFTs with more couplings between
operators as discussed in \cite{hofman, Myers2}. As a matter of
fact, the holographic study of the quasi-topological gravity has
been carried out in many
references\cite{Myers2,amsel,Kuang,Fadafan,Siani} and its dual
CFTs display much richer structures and novel features. Specially,
in this theory two central charges relating to the conformal
anomaly can be unequal, which brings in a non-zero but much lower
bound of the ratio of shear viscosity to density entropy and hence
violation of the Kovtun-Son-Starinet (KSS) bound\cite{Myers2}. In
this paper, we intend to continue our previous investigation on
holographic s-wave superconductors in \cite{Kuang}, and construct
a holographic p-wave model in the framework of quasi-topological
gravity. We are specially concerned with the anisotropic behavior
of the gauge fields and intend to compare our results with other
p-wave models in Einstein's and Gauss-Bonnet gravity. We will also
discuss the charge transport using a linear response theory, with
a special interest in its behavior in the low frequency limit.

We organize our paper as follows. In Sec.II, we present the
holographic setup for a p-wave superconductor in quasi-topological
gravity, then study the superconducting phase transition in the
probe limit, focusing on the variation of the critical temperature
with the coupling parameters. Sec.III contributes to the numerical
evaluation of the anisotropic conductivity. Based on the data
obtained we mainly discuss the following two issues. One is on the
change of the ratio of the frequency gap to the critical temperature
with the frequency, and the other is on the low frequency behavior
of the conductivity. Discussions and conclusions are given in
Sec.IV.

\section {quasi-topological holographic p-wave superconductors}
We start with the five-dimensional quasi-topological gravity
 with an SU(2) Yang-Mills gauge field. The bulk action is
given as
\begin{equation}\label{a}
S_{bulk}=\int d^5x\sqrt{-g}\Big[\frac{1}{16\pi
G_5}(R+\frac{12}{L^2}+\frac{\alpha L^2}{2}\mathcal{X}_4+\frac{7\beta
L^4}{8}\mathcal
{Z}_5)-\frac{1}{4g_{YM}^2}(F^i_{\mu\nu}F^{{i\mu\nu}})\Big],
\end{equation}
where $G_5$ is the Newton constant in five-dimensional theory, and
$\alpha$, $\beta$ and $g_{YM}$ are Gauss-Bonnet coupling parameter,
curvature-cubed interaction parameter and Yang-Mills coupling
parameter, respectively. $F^i_{\mu\nu}$ is the field strength of
 Yang-Mills gauge field with SU(2) gauge symmetry and $i$ is the internal index. Here
$\mathcal{X}_4$ is the Gauss-Bonnet term
\begin{equation}\label{b}
\mathcal{X}_4=R_{\mu\nu\rho\sigma}R^{\mu\nu\rho\sigma}-4R_{\mu\nu}R^{\mu\nu}+R^2,
\end{equation}
and $\mathcal{Z}_5$ is a curvature-cubed term with the form
\begin{eqnarray}\label{c}
\mathcal{Z}_5={R_{\mu\nu}}^{\rho\sigma}{R_{\rho\sigma}}^{\alpha\beta}{R_{\alpha
\beta}}^{\mu\nu}+\frac{1}{14}(21R_{\mu\nu\rho\sigma}{R^{\mu\nu\rho\sigma}}R
-120R_{\mu\nu\rho\sigma}{R^{\mu\nu\rho}}_{\alpha}R^{\sigma\alpha}\nonumber\\
+144R_{\mu\nu\rho\sigma}R^{\mu\rho}R^{\nu\sigma}
+128{R_\mu}^{\nu}{R_\nu}^{\rho}{R_\rho}^{\mu}-108{R_\mu}^{\nu}{R_\nu}^{\mu}R+11R^3).
\end{eqnarray}
It is worthy to point out that in contrast to higher order terms
in Lovelock gravity\cite{Lovelock}, the cubed terms above are not
just topological but have contributions to equations of motion for
bulk fields. Through this paper we will only take account of the
probe limit of the theory. Namely, we will neglect the back
reaction of the Yang-Mills field on the background metric in the
large $N_{c}$ limit, where $N_{c}$ is the number of degrees of
freedom per point in the dual free field theory \footnote{The
large $ N_{c}$ limit implies $G_5\sim N_c^{-2} \rightarrow0$,
leading to a decoupling between the matter field with a finite
$g_{YM}$ and the gravity in the action $S_{bulk}$.}.

In this limit stable AdS black hole solutions in five-dimensional
spacetime have been found in \cite{Myers} and they can be described
as
\begin{equation}\label{d}
ds^2=\frac{r^2}{L^2}(-N(r)^2f(r)dt^2+dx^2+dy^2+dw^2)+\frac{L^2}{r^2f(r)}dr^2,
\end{equation}
where $f(r)$ has three different solutions for different regions in
parameter space
\begin{eqnarray}
f_1(r)&=&u+v-\frac{\alpha}{3\beta},\label{H}\\
f_2(r)&=&-\frac{1}{2}(u+v)+i\frac{\sqrt{3}}{2}(u-v)-\frac{\alpha}{3\beta},
\label{G}\\
f_3(r)&=&-\frac{1}{2}(u+v)-i\frac{\sqrt{3}}{2}(u-v)-\frac{\alpha}{3\beta},
\end{eqnarray}
with
\begin{eqnarray}\label{f}
u&=&(q+\sqrt{q^2-p^3})^{1/3},\hspace{1cm}v=(q-\sqrt{q^2-p^3})^{1/3},\label{n}\nonumber\\
p&=&\frac{3\beta+\alpha^2}{9\beta^2},\hspace{2.7cm}q=-\frac{2\alpha^3+9\alpha\beta+27\beta^2(1-\frac{r_H^4}{r^4})}{54\beta^3}.
\end{eqnarray}
$L$ is the AdS radius and
$N(r)=N=1/\sqrt{f(r)\mid_{r\rightarrow\infty}}$ is the lapse
function\footnote{In order to get a normalized velocity of light on
the boundary, the lapse function $N$ usually should not be set to
unit. This gauge is different from that in \cite{cai2} and some
discrepancy of our results for GB gravity with those in \cite{cai2}
can be ascribed to this different gauge. Moreover, the argument that
$N(r)$ is a constant can be seen in \cite{Myers}.}. Now it is
straightforward to obtain the Hawing temperature of these black
holes, which is
\begin{equation}\label{g}
T=\frac{N}{4\pi}f'(r)|_{r=r_H}=\frac{Nr_H}{\pi L^2}.
\end{equation}
It will also be viewed as the temperature of the dual CFT on the
boundary.

Now we turn to construct the holographic p-wave superconductors.
Firstly we need to solve the Yang-Mills equations in a fixed black
hole background. Following the strategy presented in \cite{Gubser},
we take the ansatz as follows
\begin{equation}\label{h}
A_a=A^{i}_\mu
\tau^i(dx^\mu)_a=\tilde{\phi}(r)\tau^3(dt)_a+\tilde{\psi}(r)\tau^1(dx)_a,
\end{equation}
where $\tau^{i}=\sigma^{i}/2i$ (i=1,2,3) with commutation relations
$[\tau^i,\tau^j]=i\epsilon^{ijk}\tau^k$ are SU(2) generators.
 In (\ref{h}) the nonvanishing $\tilde{\psi}(r)$ breaks the $U(1)_{3}$
gauge symmetry generated by $\tau^3$. We may interpret it as the
p-wave superconducting phase transition from the side of CFT on the
boundary, since in the dual field theory the global $U(1)_{3}$
symmetry is broken and superconducting charges can be created from
the new vacuum which corresponds to the formation of the Cooper
pairs. For convenience, we absorb the gauge couplings into the
rescaling of the gauge fields
\begin{equation}\label{i}
\phi=g_{YM}L^2\tilde{\phi},\hspace{1cm}\psi=g_{YM}L^2\tilde{\psi}.
\end{equation}
Moreover, we redefine the coordinate $z=\frac{r_H}{r}=\frac{1}{r}$
such that the position of the horizon is fixed at $z=1$, while the
boundary is $z\rightarrow 0$. Then with the ansatz in Eq.(\ref{h})
the
 equations for Yang-Mills field reduce to the following form
\begin{equation}\label{j}
\phi''-\frac{\phi'}{z}-\frac{L^2\psi^2}{z^2g}\phi=0,
\end{equation}
\begin{equation}\label{k}
\psi''+(\frac{g'}{g}+\frac{1}{z})\psi'+\frac{\phi^2}{N^2g^2z^4}\psi=0,
\end{equation}
where $g=\frac{r^2f(r)}{L^2}$ and the prime denotes a derivative
with respect to $z$. Before solving these two equations we give the
boundary conditions near the horizon and near the AdS  boundary as
follows:

\noindent $\blacklozenge$ The regularity condition at the horizon
($z=1$) requires the gauge fields should be expanded as
\begin{eqnarray}\label{l}
\psi&=&\psi^{(0)}_H+\psi^{(2)}_H(1-z)^2+\cdots\nonumber\\
\phi&=&\phi^{(1)}_H(1-z)+\cdots.
\end{eqnarray}
\noindent $\blacklozenge$ Near the AdS boundary ($z\rightarrow0$),
the asymptotical behavior of fields are like
\begin{eqnarray}\label{m}
\psi&=&\psi^{(0)}+\psi^{(2)}z^2+\cdots\nonumber\\
\phi&=&\mu+\rho z^2+\cdots.
\end{eqnarray}
In AdS/CFT dictionary, $\mu$ is the chemical potential on the
boundary while $\rho_t=2\rho$ and $\rho_n=\phi^{(1)}_H$ are
understood as the total charge density and the charge density in
the normal state, respectively. So the p-wave superconducting
charge density is $\rho_s=\rho_t-\rho_n$. A nonzero $\psi^{(0)}$
and nonzero $\psi^{(2)}$ correspond to a source and the
expectation value of vacuum for the current operator $J_x^1$ that
is dual to the gauge field $A_x^1=\psi$ respectively, so we have
\begin{equation}\label{n}
\langle J_x^1\rangle=\psi^{(2)}.
\end{equation}
For normalizable modes, the expectation value of vacuum can be
obtained by setting $\psi^{(0)}=0$.

\begin{figure} \center{
\includegraphics[scale=0.3]{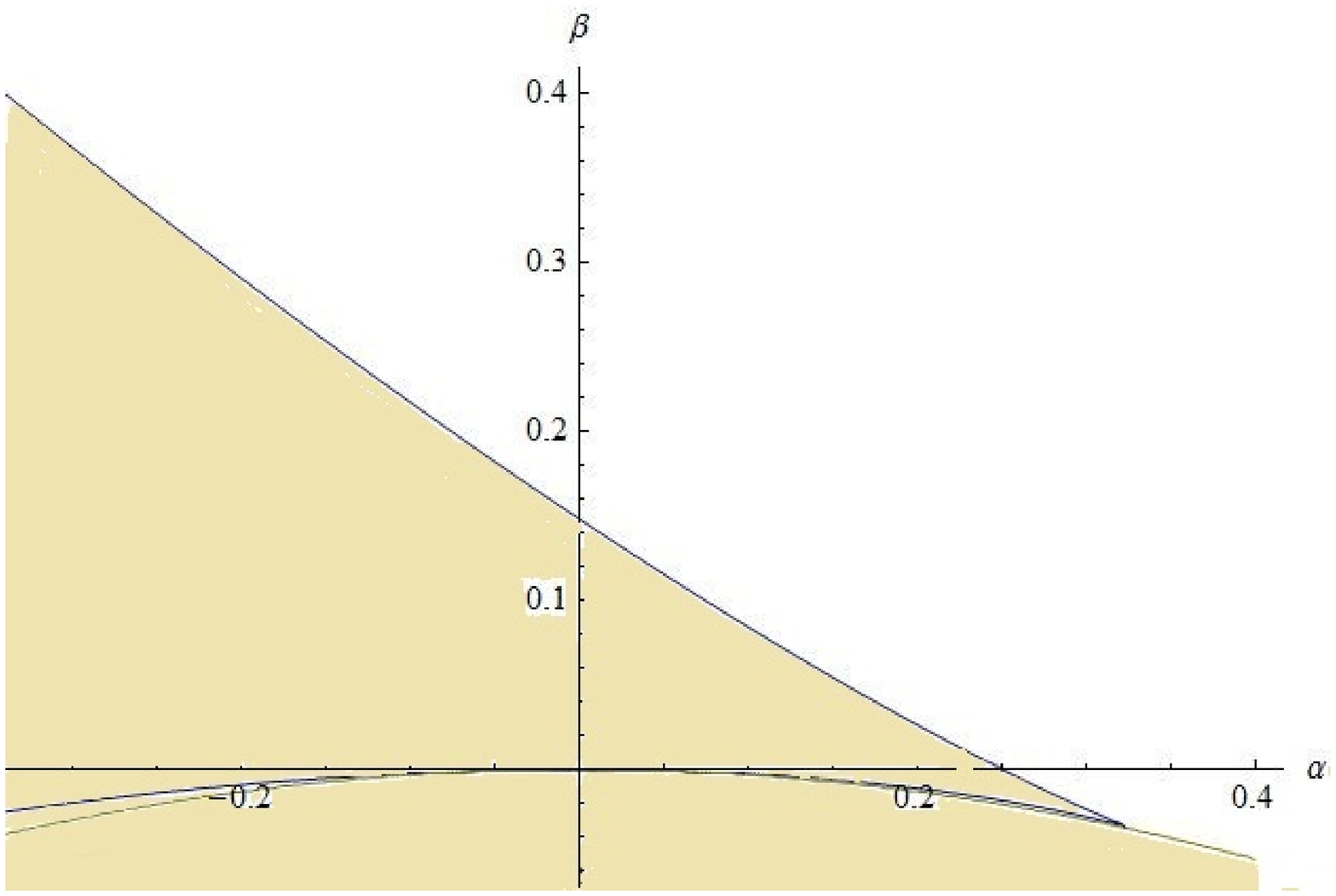}
\includegraphics[scale=0.6]{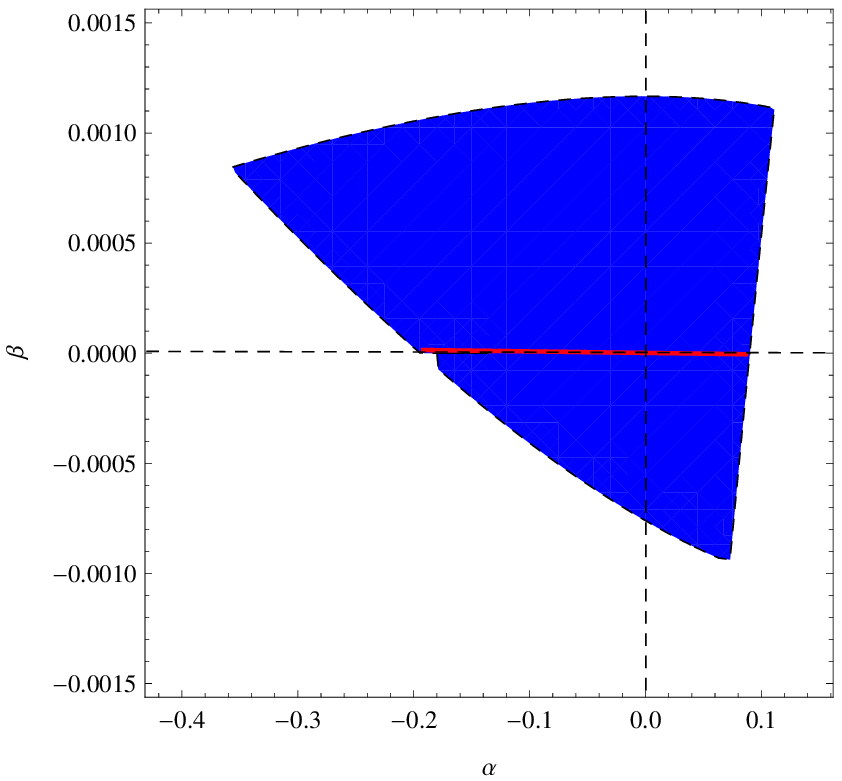}\hspace{0.0cm}
\caption{
\label{fa} The allowed range of the values of coupling parameters
$\alpha$ and $\beta$: In the left plot, the yellow region is for
the stable solution and the different regions divided by the color
lines correspond to different stable black hole solutions given by
$(5)-(7)$ (see \cite{Myers} for details); In the right plot, the blue
region is the valid parameter range restricted by consistency of the dual CFT,
and the red line corresponds to $\beta=0$, i.e., the constraint of $\alpha$ in Gauss-Bonnet gravity.}}
\end{figure}

Before doing the numerical analysis, we present some remarks on
the allowed range of the values of the coupling parameters
$\alpha$ and $\beta$. First of all, to obtain the stable black
hole solutions without ghost modes or naked singularity, one finds
that the allowed parameter range is constrained in the region as
illustrated in the left plot of FIG.\ref{fa}\cite{Myers}. On the
other hand, to ensure the positivity of central charges, energy
flux and a well-defined causality for the dual CFT,  $\alpha$ and
$\beta$ are further severely confined into a small region as
dictated in the right plot of FIG.\ref{fa}, which has originally
been presented in \cite{Myers2}. Thus in our paper we will
investigate the holographic superconductivity with parameter
values restricted in this small region. However, in this region
the coupling constant $\beta$ is severely restricted in a narrow
interval, roughly from $\beta=-0.001$ to $\beta=0.001$.
Graphically when one changes the value of $\beta$ in this region,
the plotting is probably not sensitive enough to illustrate the
changes of the physical quantities with different values of the
coupling constants. Thus we take two actions in our plotting. One
is to enlarge a local region in figures to demonstrate the
shifting tendency of the curves with the values of coupling
parameters. Secondly, we also take some value for $\beta$ from the
region in the left plot of FIG.1 for comparison, for instance
$\beta=0.1$. In the probe limit we are allowed to do this since
the back reaction of the perturbations is not taken into account
and we may still obtain the stable phase in dual field theory, but
it is warned that the curves obtained for these values are
potentially instable for a system when the back reaction of the
perturbations is considered, thus should not be trusted seriously.

\begin{figure}
\center{
\includegraphics[scale=0.4]{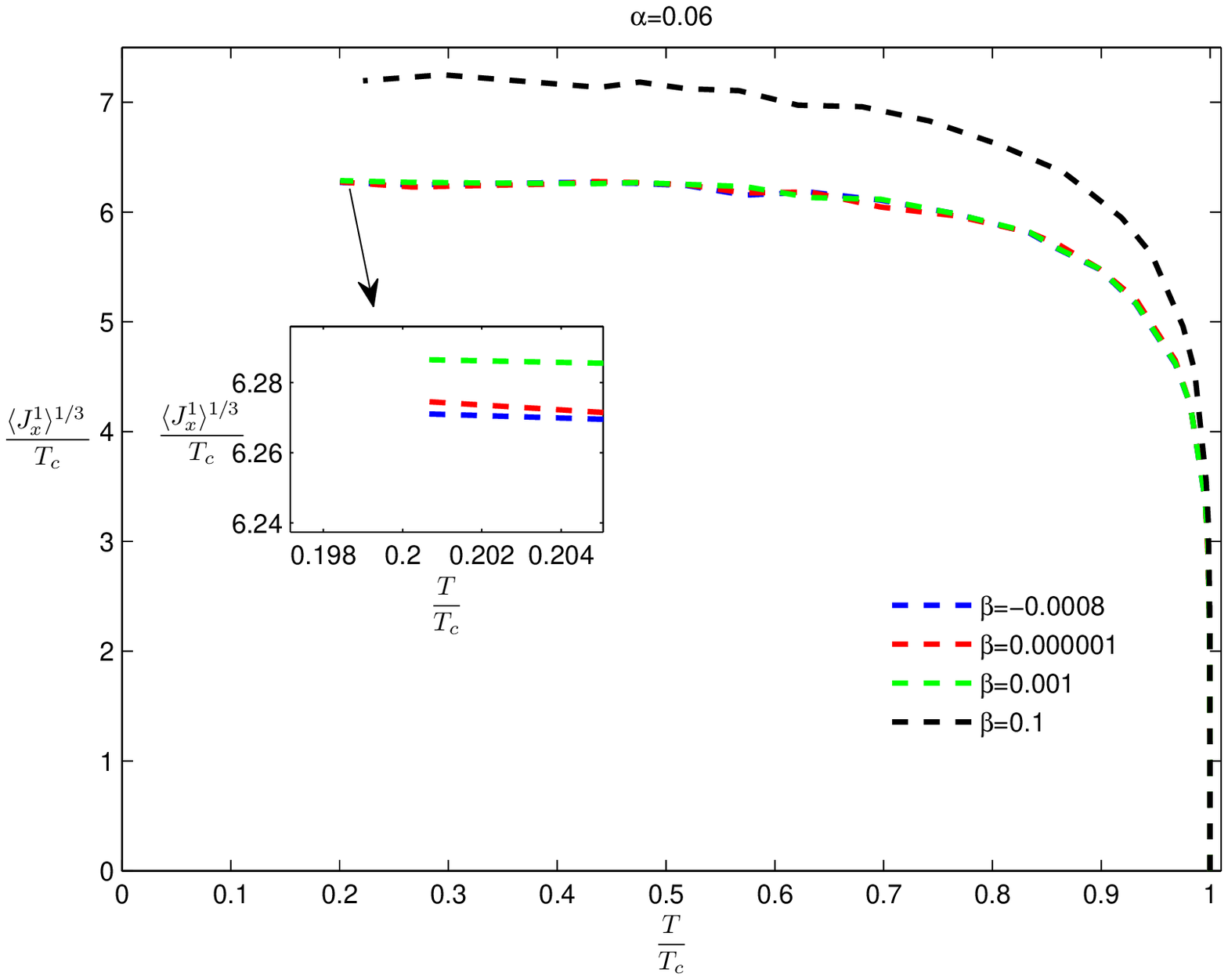}
\includegraphics[scale=0.4]{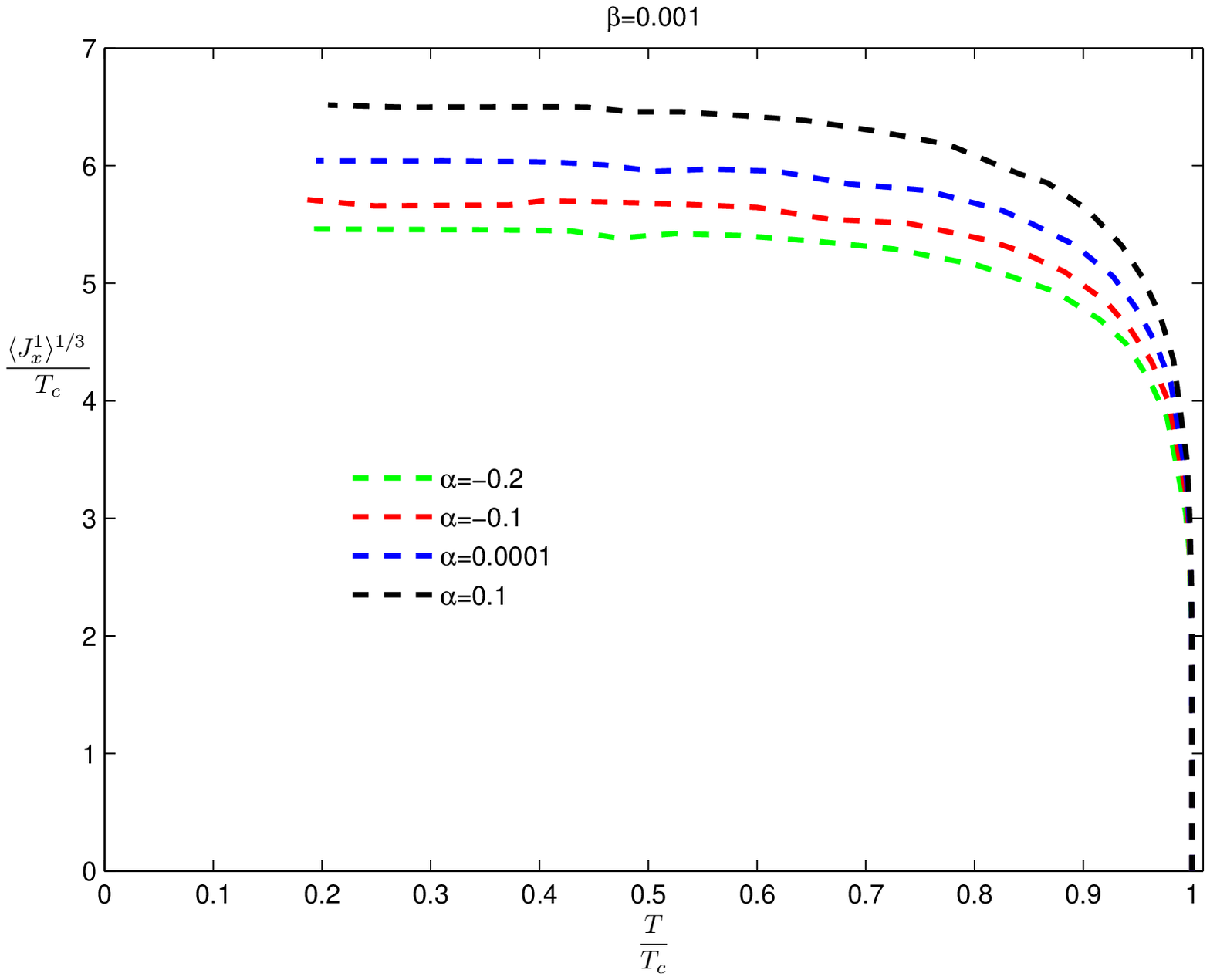}
\caption{\label{fb}The condensate as a function of the temperature with
different values of coupling parameters. In the left figure, the
Gauss-Bonnet parameter $\alpha$ is fixed at $0.06$. $\beta$ is
fixed at $0.001$ in the right figure. In both cases the condensation
tends to increase with the coupling parameters.}}
\end{figure}

 Now to explore the p-wave superconducting phase, we need find
nonzero solutions for $\psi$ by numerical analysis. In our
program, we set $L=1$ and find the numerical solutions to the
differential EOMS from the horizon to the AdS boundary, namely,
from $z\rightarrow1$ to $z\rightarrow0$.  In Figure \ref{fb},
we illustrate the condensation of $J_x^1$ as the function of the
temperature with different values of coupling parameters $\alpha$
and $\beta$. Note that different from the case of s-wave
superconductor, here we plot the dimensionless quantity
$\frac{\sqrt[3]{J_x^1}}{T_c}$ with respect to $\frac{T}{T_c}$ since
the conformal dimension of $J_x^1$ is 3 rather than 1 for $T_c$.
 FIG.2 indicates that the order parameter has the behavior
$\langle J_x^1\rangle\propto(1-T/T_c)^{1/2}$ near the critical
temperature, and the value of $\frac{\sqrt[3]{J_x^1}}{T_c}$
increases with both the Gauss-Bonnet coupling parameter $\alpha$ and
the curvature-cubed coupling parameter $\beta$, which is similar to
the phenomenon obtained in the s-wave superconductor in
quasi-topological gravity\cite{Kuang}, as well as the p-wave
superconductors in Gauss-Bonnet gravity \cite{cai2}.

\begin{figure}
\center{
\includegraphics[scale=0.4]{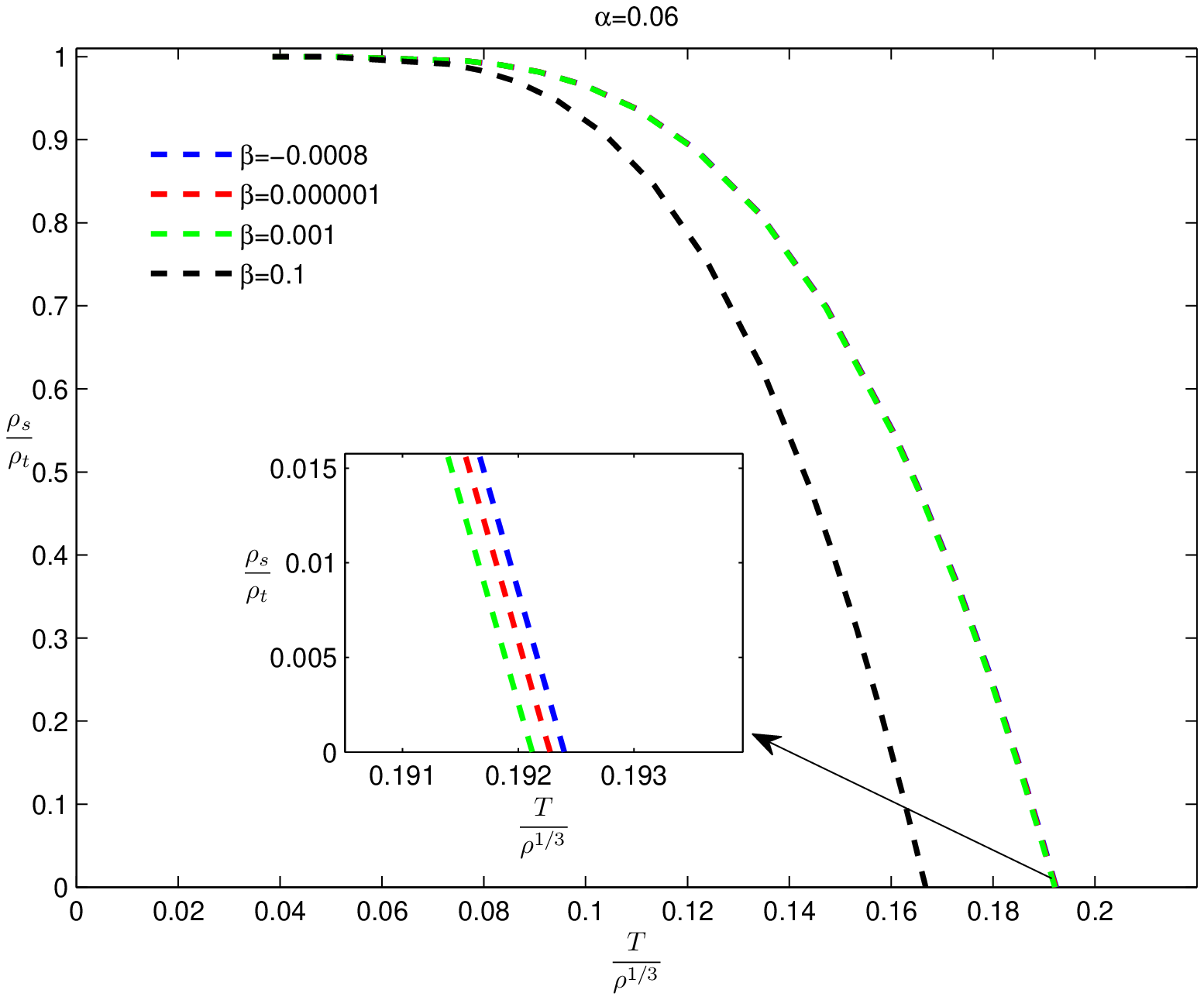}
\includegraphics[scale=0.4]{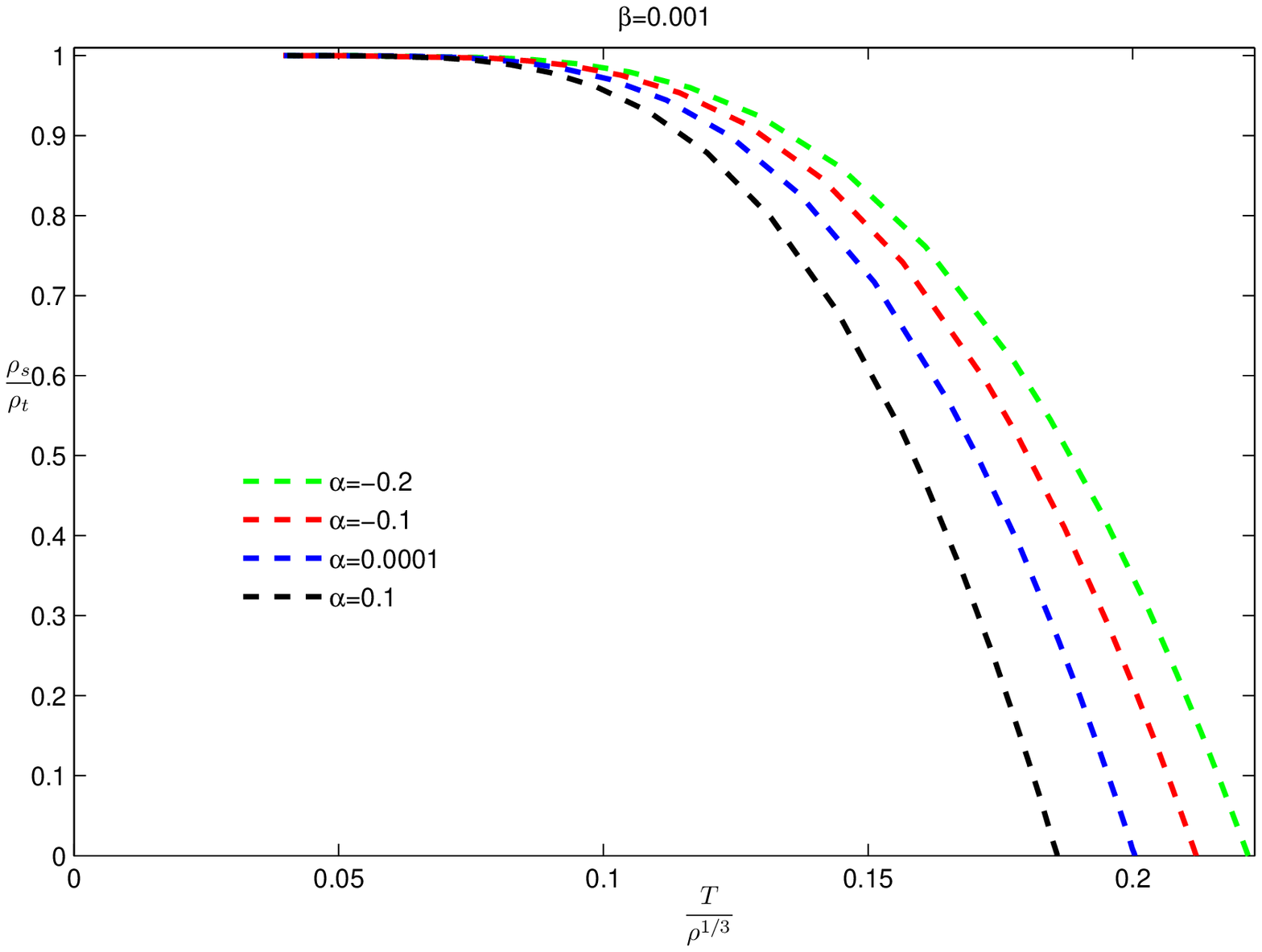}
\caption{\label{fc}The ratio of the superconducting charge density to the
total charge density v.s. the temperature with different values of
coupling parameters.}}
\end{figure}

In figure \ref{fc}, we plot $\rho_s/\rho_t$ v.s. $\frac{T}{\sqrt[3]{\rho}}$
by changing either of the coupling parameters $\alpha$ and $\beta$.
The critical temperatures for different coupling parameters can be
read off from the intersects with the horizontal axis and their
values are listed in TABLE I.
 \begin{table}
\begin{center}
\begin{tabular}{| l | c | r | }
\hline$\alpha$ &$\beta$ &$T_c$ \\
\hline0.06 & -0.0008 & 0.1924$\rho^{1/3}$ \\
\hline 0.06 &0.000001&0.1923$\rho^{1/3}$\\
\hline 0.06 & 0.001 &0.1921$\rho^{1/3}$ \\
\hline 0.06& 0.1 & 0.1667$\rho^{1/3}$ \\
\hline-0.2& 0.001&0.2218$\rho^{1/3}$\\
\hline -0.1& 0.001&0.2120$\rho^{1/3}$\\
\hline 0.0001& 0.001 &0.2004$\rho^{1/3}$ \\
\hline 0.1 & 0.001 & 0.1858$\rho^{1/3}$ \\
\hline
\end{tabular}
\caption{The change of the critical temperature with the coupling
parameter $\alpha$ and $\beta$. The critical temperatures
corresponding to the parameter values in FIG.\ref{fc} are listed in two
tables respectively.}
\end{center}
\end{table}
\begin{figure} \center{
\includegraphics[scale=0.7]{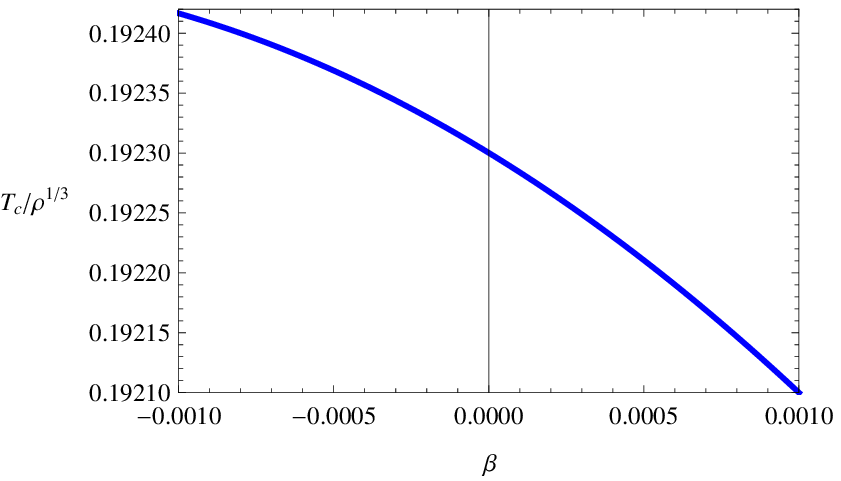}\hspace{2.0cm}
\includegraphics[scale=0.7]{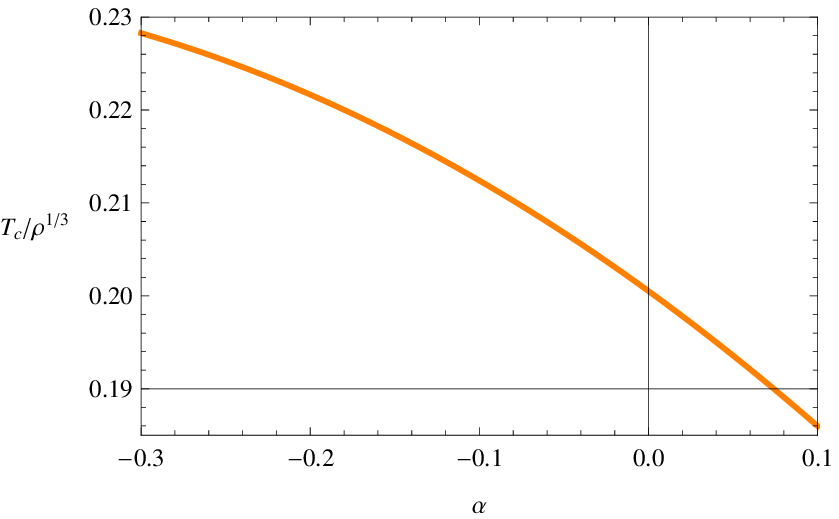}\hspace{0.0cm}
\caption{\label{fd}The relations between $T_c$ and the coupling parameters
$\alpha$ and $\beta$, respectively. In the left figure the line is
for $\alpha=0.06$, while in the right figure the line is for
$\beta=0.001$. }}
\end{figure}
We can see that the critical temperature decrease as either
$\alpha$ or $\beta$ increases, implying that with the increase of
the curvature-cubed term or the Gauss-Bonnet term the occurrence
of condensation should become harder. This tendency is the same as those in the previous papers\cite{Kuang,cai2}. For
explicitness we demonstrate the dependence of the critical
temperature on either of the coupling parameter in FIG.\ref{fd}.
Moreover, in \cite{Myers2} we know the conformal field dual to the
quasi-topological gravity is characterized by central charges, $c$
and $a$, and flux parameters, $t2$ and $t4$. Explicitly, these
parameters can be related to the coupling parameters in the bulk
as follows
\begin{eqnarray}\label{aaa}
\delta&=&\frac{c-a}{c}=\frac{4f_{\infty}(\alpha-3\beta f
_{\infty})}{1-2\alpha f_{\infty}-3\beta f_{\infty}^{2}},
\end{eqnarray}
\begin{eqnarray}
t2&=&\frac{24f_{\infty}(\alpha-87f_{\infty}\beta)}{1-2\alpha
f_{\infty}-3\beta f_{\infty}^{2}}\\
t4&=&\frac{3780f_{\infty}^{2}\beta}{1-2\alpha f_{\infty}-3\beta
f_{\infty}^{2}},\label{aab}
\end{eqnarray}
where $ f_{\infty}$ satisfies $1-f_{\infty}+\alpha
f_{\infty}^{2}+\beta f_{\infty}^{3}=0$. Among these parameters any
two of them is enough to describe the correlations in the dual
CFT. For example, we can calculate $\delta$ and $t_4$ as the free
parameters and fit how $T_c$ changes with the changing of $\delta$
and $t4$ respectively. However, as discussed in \cite{Kuang}, a
clear rule from the CFT side is still missing, thus we will not
show the fitting plot of the critical temperature versus $\delta$
or $t4$ in our current paper.

In the next section, we turn to investigate the conductivity and
find its new characters comparing with the s-wave superconductors.

\section{conductivity}
In this section, we will study the charge transport and linear
response of the boundary system. In the linear response theory, a
central quantity is the retarded Green function. According to the
AdS/CFT dictionary, if we want to know the retarded Green function
of the $U(1)$ current, we just need to study the propagation of the
linear perturbation of the gauge field in the bulk in the probe
limit. So, in the following, we are interested in the linear
response of the $\tau^3$ component of the Yang-Mills gauge field. As
discussed in \cite{Gubser}, we can consider an alternating
current(AC) on the boundary by introducing a time-dependent
perturbation for the gauge field
\begin{equation}\label{0}
A \rightarrow A+\delta A,
\end{equation}
where
\begin{eqnarray}\label{p}
\delta A=e^{-i\omega
t}\Big[\Big(a_t^1(r)\tau^1+a_t^2(r)\tau^2\Big)dt+a_x^3(r)\tau^3dx+a_y^3(r)\tau^3dy\Big].
\end{eqnarray}
Though the condensation of $\psi$ breaks the rotational $SO(3)$
symmetry associated with $x$ direction, the system still has an
$SO(2)$ symmetry in $y-w$ plane. Hereafter, we neglect the
perturbation along $w$-axis and only consider the electrical
conductivity $\sigma_{xx}$ and $\sigma_{yy}$.

\section*{ $1\centerdot$ $\sigma_{yy}$ component}
Substituting the ansatz in (\ref{p}) into the Yang-Mills equations,
we obtain the equation of motion for $a_y^3$
\begin{equation}\label{r}
{a^3_y}''+(\frac{g'}{g}+\frac{1}{z}){a^3_y}'+\Big(\frac{\omega^2}{N^2g^2z^4}-\frac{L^2\psi^2}{z^2g}\Big)a^3_y=0,
\end{equation}
where the prime is relative to $z=1/r$. In general, different modes
of the perturbations are mixed, however, from the equation above we
notice that $a^3_y$ is actually decoupled from other components. In
addition, the equation is similar to that for s-wave superconductors
\cite{Kuang,cai2}, so we can obtain the conductivity $\sigma_{yy}$
in a parallel way. Specifically, we choose the ingoing wave
condition for $a^3_y$ near the horizon, then we have
\begin{equation}\label{s}
a_y^3=(1-z)^{\frac{-i\omega}{4N}}[1+{a_y^3}^{(1)}(1-z)+{a_y^3}^{(2)}(1-z)^2+\cdots].
\end{equation}
Moreover, the behavior of $a^3_y$ in the asymptotical AdS boundary
($z\rightarrow0$) is
\begin{eqnarray} \label{t}
a^3_y=a_y^{3(0)}+a_y^{3(2)}z^2+\frac{a_y^{3(0)}\omega^2N^4}{2}
(\log\Lambda/z)z^2,
\end{eqnarray}
where $a_y^{3(0)}$, $a_y^{3(2)}$ and $\Lambda$ are integration
constants. The last term in (\ref{t}) will lead to a divergence when
one calculates the Green function, however, such a logarithmic
divergence can be canceled with a boundary counterterm in the
renormalization procedure\cite{Robinson}. Thanks to the standard
AdS/CFT dictionary, the conductivity can be expressed through the
retarded Green function as follows\cite{Son}
\begin{eqnarray}\label{u}
  \sigma (\omega) = \frac{1}{i \omega} G^R(\omega)\Big|_{\textbf{k=0}}=-\frac{1}{i \omega}\lim_{r\rightarrow\infty}N g(r)ra^3_y{a^3_y}',
\end{eqnarray}
Thus we find the conductivity $\sigma_{yy}$ is\footnote{As pointed
out in \cite{Barclay}, the term $-i\omega N lnN$ should not be
neglected. Though we take a different gauge here, the asymptotic
analysis near the boundary is similar and a direct calculation shows
that the result is the same.}
\begin{eqnarray}\label{w}
\sigma_{yy}=&=&\frac{2a_y^{3(2)}}{i\omega N^3a_y^{3(0)}}-i\omega N lnN +\frac{iN\omega}{2} \ .
\end{eqnarray}

\begin{figure}
\center{
\includegraphics[scale=0.4]{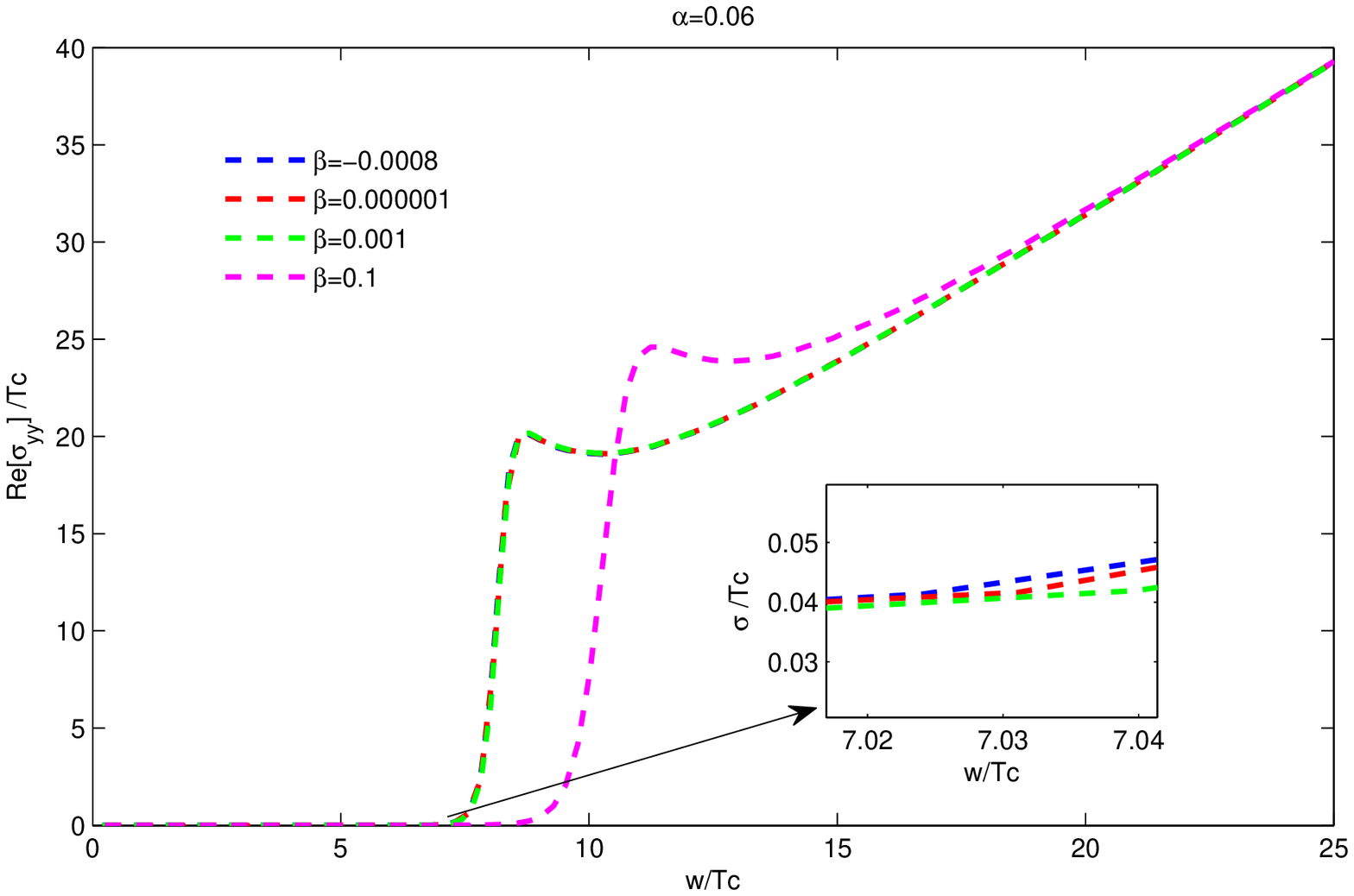}\hspace{0.0cm}
\includegraphics[scale=0.4]{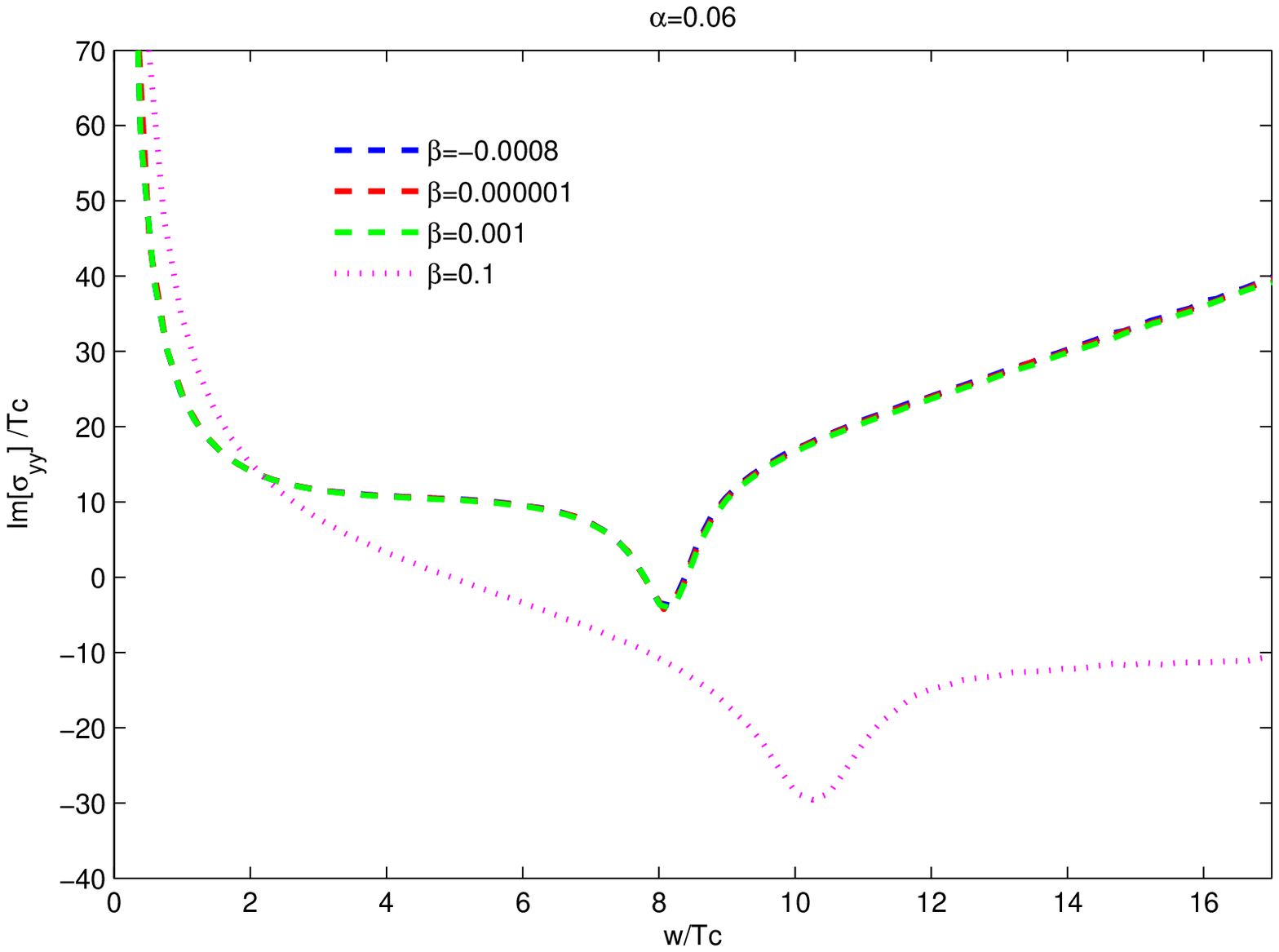}\\ \hspace{0.0cm}
\caption{\label{fe}The conductivity $\sigma_{yy}$ for the p-wave
superconductors with a fixed Gauss-Bonnet parameter $\alpha=0.06$. The
left figure is for the real part of $\sigma_{yy}$ while the right
one is for the imaginary part of $\sigma_{yy}$.}}
\end{figure}

\begin{figure}
\center{
\includegraphics[scale=0.4]{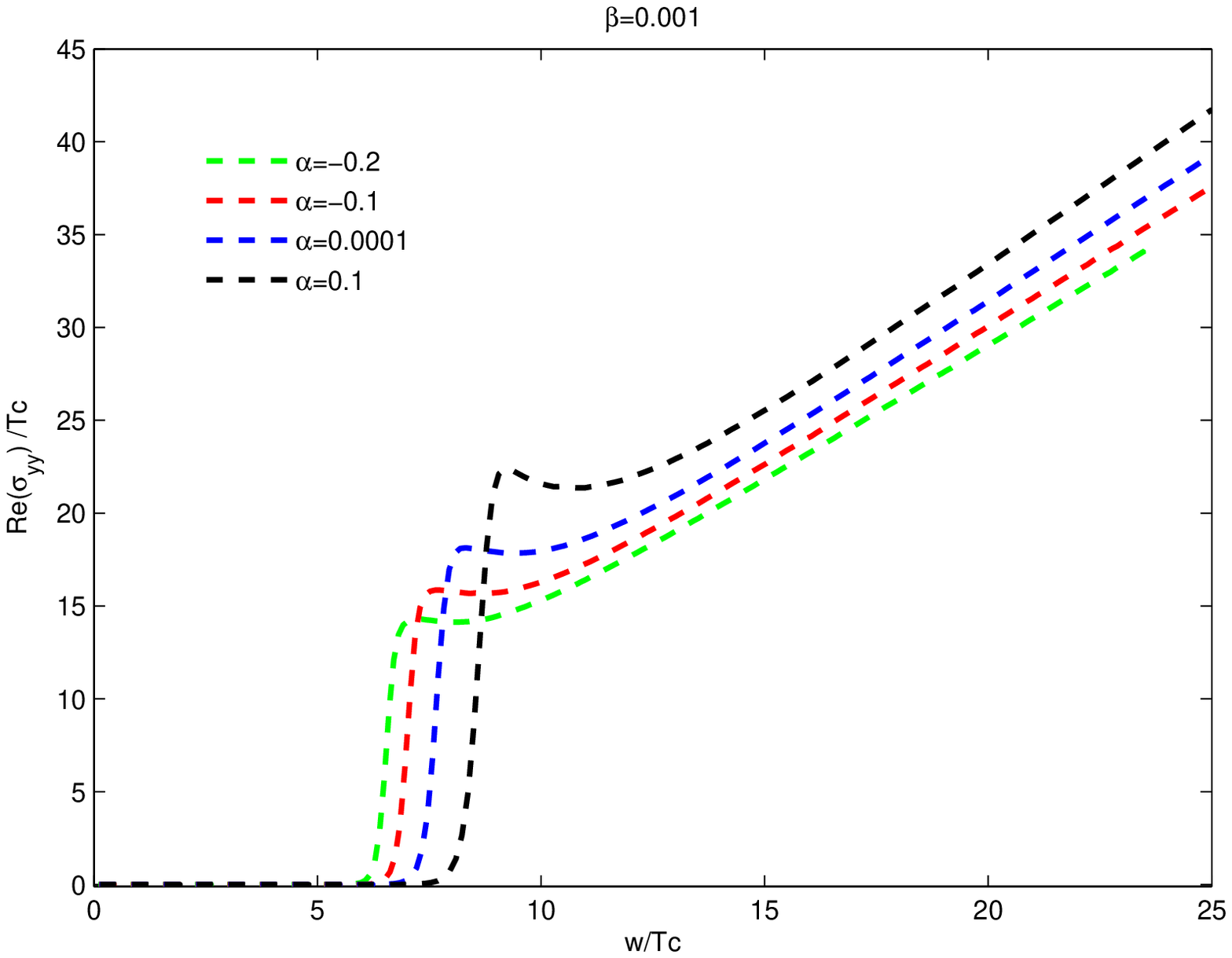}\hspace{0.0cm}
\includegraphics[scale=0.4]{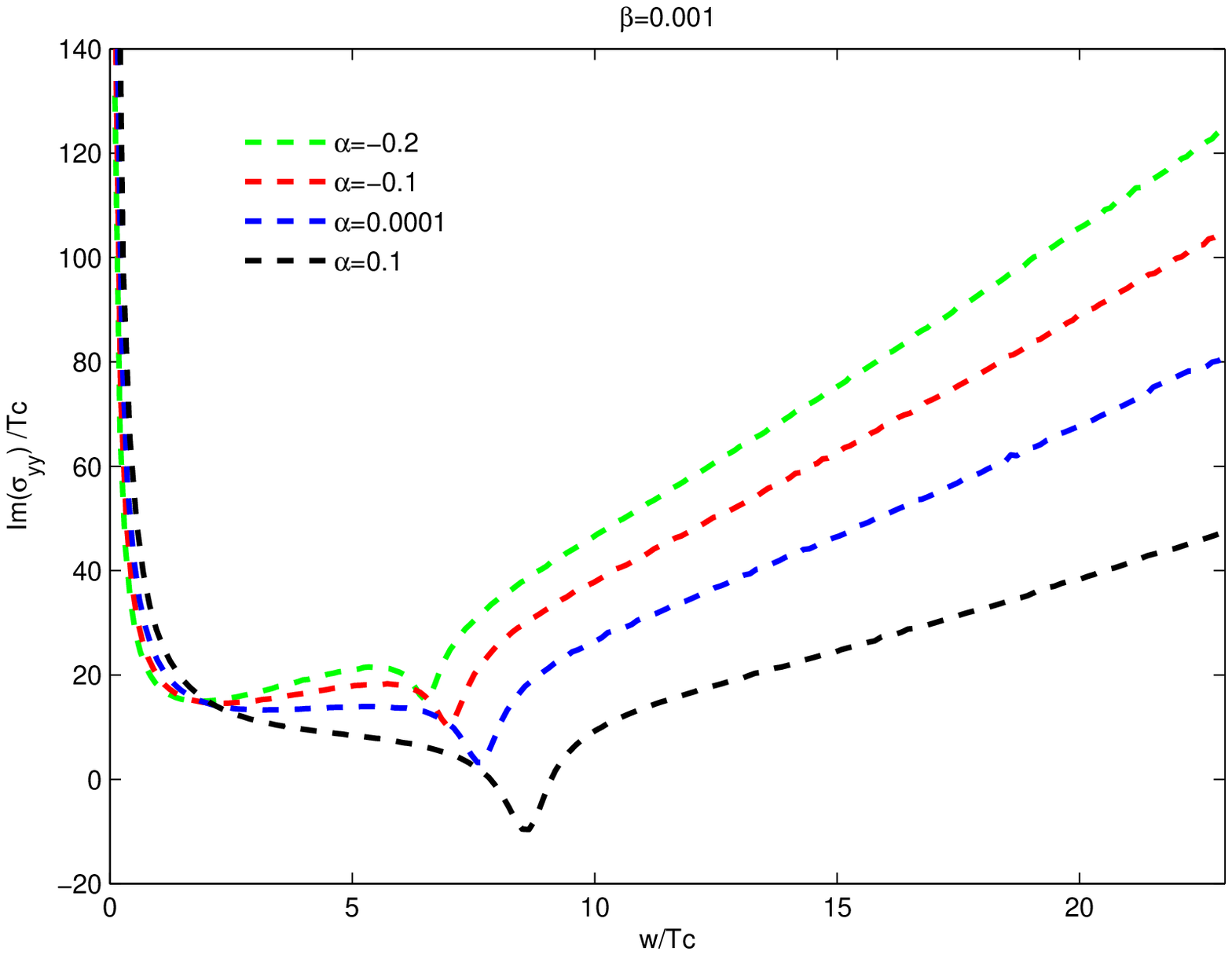}\hspace{0.0cm}
\caption{\label{ff}The conductivity $\sigma_{yy}$ for the p-wave
superconductors with a fixed the curvature-cubed parameter
$\beta=0.001$. Similarly, the left figure is for the real part of
$\sigma_{yy}$ while the right one is for the imaginary part of
$\sigma_{yy}$.}}
\end{figure}

Given the boundary conditions (\ref{s}) and (\ref{t}), we
numerically solve equation (\ref{r}) to obtain $a_y^{3(0)}$ and
$a_y^{3(2)}$. As a result, the dependent relation between
$\sigma_{yy}$ and $\omega$ for different coupling parameters $\beta$
and $\alpha$ are shown in FIG.\ref{fe} and FIG.\ref{ff}, respectively.
From these two figures, it is clear that when the frequency
$\omega$ is large enough, the real part of $\sigma_{yy}$ always
increases while the imaginary part becomes linear due to the
dominant term $\omega N lnN +\frac{N\omega}{2}$. In addition, from
the imaginary part of conductivity one finds that the ratio
$\omega_g/T_c$ increases with either $\alpha$ or $\beta$. This
behavior is much similar to that of $\sigma$ in the s-wave
case\cite{Kuang}, while it is interesting to note that for
p-wave superconductor the values of the real part of conductivity
increase with the coupling parameters. Since this quantity
corresponds to the imaginary part of the Green function which
characterizes the dissipation of the charge transport, this
phenomenon is analogous to what happens in holographic
hydrodynamics as described in figure 2 in \cite{Myers2}.

\section*{ $2\centerdot$ $\sigma_{xx}$ component}
Now we intend to study the component $\sigma_{xx}$, which can be
obtained by solving the equation of motion for $a_x^3$ in the bulk.
In contrast to $a_y^3$, since the component $a_x^3$ couples with the
fields $a_t^1$ and $a_t^2$ in the linearized Yang-Mills equations,
besides three second-order coupling equations of motion
\begin{eqnarray}\label{x}
{a^3_x}''+(\frac{1}{z}+\frac{g'}{g}){a^3_x}'+\frac{\omega^2 }{g^2 N^2 z^4}a^3_x-\frac{i\omega a_t^2+a_t^1 \phi}{g^2 N^2 z^4}\psi&=&0,\nonumber\\
{a^1_t}''-\frac{1}{z}{a^1_t}'+\frac{L^2a^3_x\phi\psi}{z^2g}&=&0,\nonumber\\
{a^2_t}''-\frac{1}{z}{a^2_t}'-\frac{L^2\psi^2}{z^2g}a^2_t-\frac{iL^2\omega
a^3_x\psi}{z^2g}&=&0,
\end{eqnarray}
we need to solve another two first-order equations together
\begin{eqnarray}\label{y}
-i\omega{a_t^1}'-\phi{a^2_t}'+a^2_t\phi'&=&0,\nonumber\\
\phi{a_t^1}'-i\omega{a_t^2}'+L^2N^2g\psi{a_x^3}'z^2-a_t^1\phi'-L^2a_x^3N^2g\psi'z^2&=&0.
\end{eqnarray}
Again using the ingoing wave condition near the horizon,
we have the following asymptotical behavior of $a_y^3$, $a_t^1$ and
$a_t^2$:
\newline
 $\blacktriangleright$ Near the horizon
($z\rightarrow1$)
\begin{equation}\label{aa}
 a_x^3=(1-z)^{\frac{-i\omega}{4N}}[1+{A_x^3}^{(1)}(1-z)+{A_x^3}^{(2)}(1-z)^2+\cdots],
 \end {equation}
\begin{equation}\label{ab}
 a_t^1=(1-z)^{\frac{-i\omega}{4N}}[{A_t^1}^{(2)}(1-z)^2+{A_t^1}^{(3)}(1-z)^3+\cdots],
 \end {equation}
\begin{equation}\label{ac}
 a_t^2=(1-z)^{\frac{-i\omega}{4N}}[{A_t^2}^{(1)}(1-z)+{A_t^2}^{(2)}(1-z)^2+\cdots].
 \end {equation}
$\blacktriangleright$ Near the boundary of the AdS bulk
($z\rightarrow0$)
\begin{equation}\label{ad}
a_x^3={a_x^3}^{(0)}+{a_x^3}^{(2)}z^2+\frac{{a_x^3}^{(0)}\omega^2N^4log(\Lambda/z)z^2}{2}+\cdots,
\end{equation}
\begin{equation}\label{ae}
a_t^1={a_t^1}^{(0)}+{a_t^1}^{(2)}z^2+\cdots,
\end{equation}
\begin{equation}\label{af}
a_t^2={a_t^2}^{(0)}+{a_t^2}^{(2)}z^2+\cdots.
\end{equation}

All the coefficients can be determined numerically. However, these
quantities are gauge dependent since we have not done any gauge
fixing to these components. As discussed in \cite{Gubser}, in order
to obtain the conductivity $\sigma_{xx}$ we need define a gauge
invariant quantity as follows

\begin{equation}\label{ag}
\hat{a_x^3}=a_x^3+\frac{i\omega L^2a_t^2+\phi
a_t^1}{\phi^2-\omega^2L^4}\psi.
\end{equation}
Then we have the asymptotical behavior of $\hat{a_x^3}$ near the AdS
boundary
\begin{eqnarray}\label{ah}
\hat{a_x^3}&=&{a_x^3}^{(0)}+{a_x^3}^{(2)}z^2+\frac{{a_x^3}^{(0)}\omega^2N^4log(\Lambda/z)z^2}{2}+\frac{i\omega
L^2{a_t^2}^{(0)}+\mu{ a_t^1}^{(0)}}{\mu^2-\omega^2L^4}{\psi}^{(2)}z^2\nonumber\\
&=&{a_x^3}^{(0)}+{\hat{a_x^3}}^{(2)}z^2+\frac{{a_x^3}^{(0)}\omega^2N^4log(\Lambda/z)z^2}{2},
\end{eqnarray}
where we have defined
\begin{equation}\label{22}
{\hat{a_x^3}}^{(2)}={a_x^3}^{(2)}+\frac{i\omega
L^2{a_t^2}^{(0)}+\mu{ a_t^1}^{(0)}}{\mu^2-\omega^2L^4}{\psi}^{(2)}.
\end{equation}
As a result, the conductivity $\sigma_{xx}$ has the form
\begin{equation}\label{19}
\sigma_{xx}=\frac{1}{i\omega}G^{R}(\omega)\mid
_{k=0}=-\frac{2i{\hat{a_x^3}}^{(2)}}{{a_x^3}^{(0)}\omega
N^3}-i\omega N lnN+\frac{1}{2}i\omega N.
\end{equation}
\begin{figure}
\center{
\includegraphics[scale=0.4]{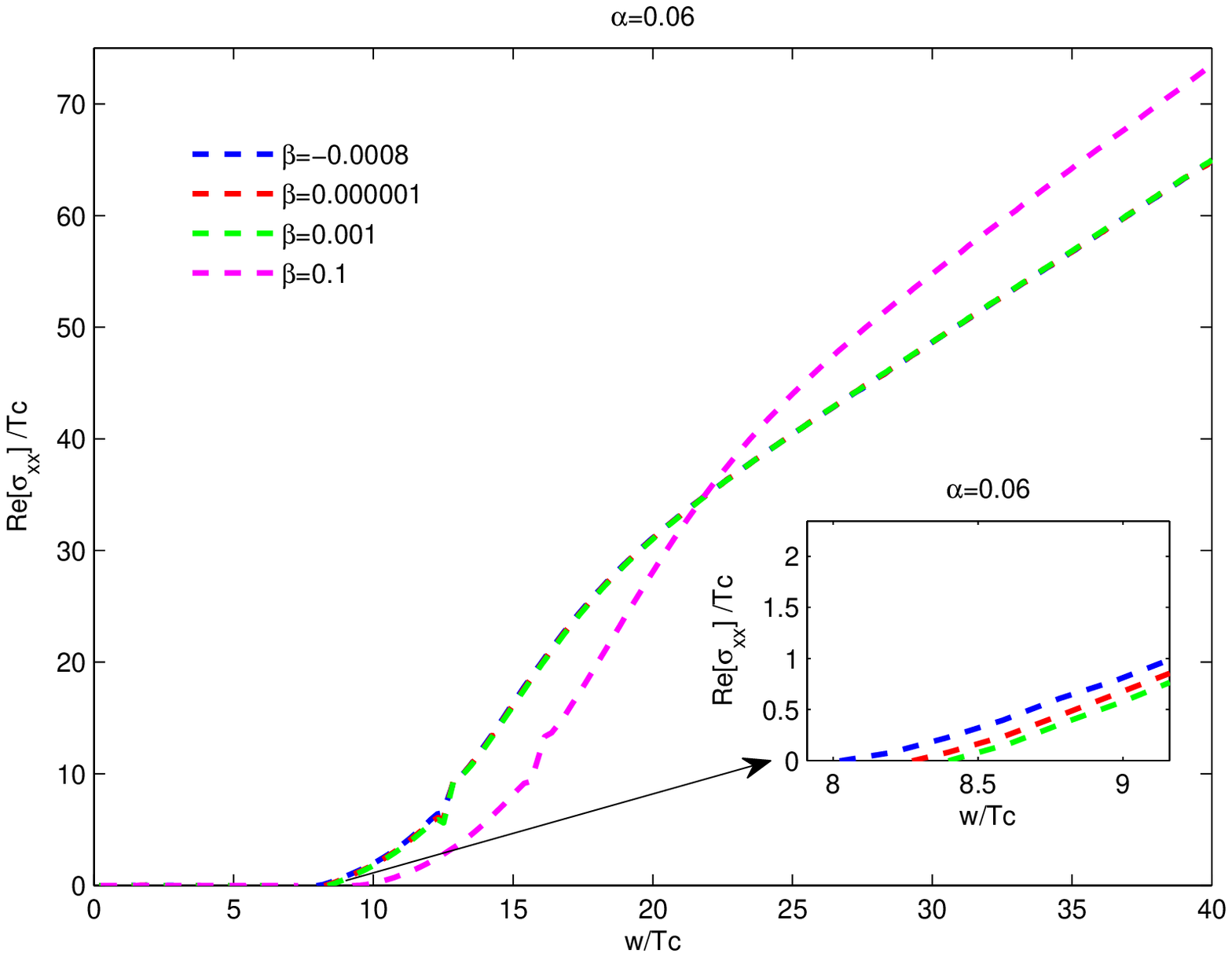}\hspace{0.0cm}
\includegraphics[scale=0.4]{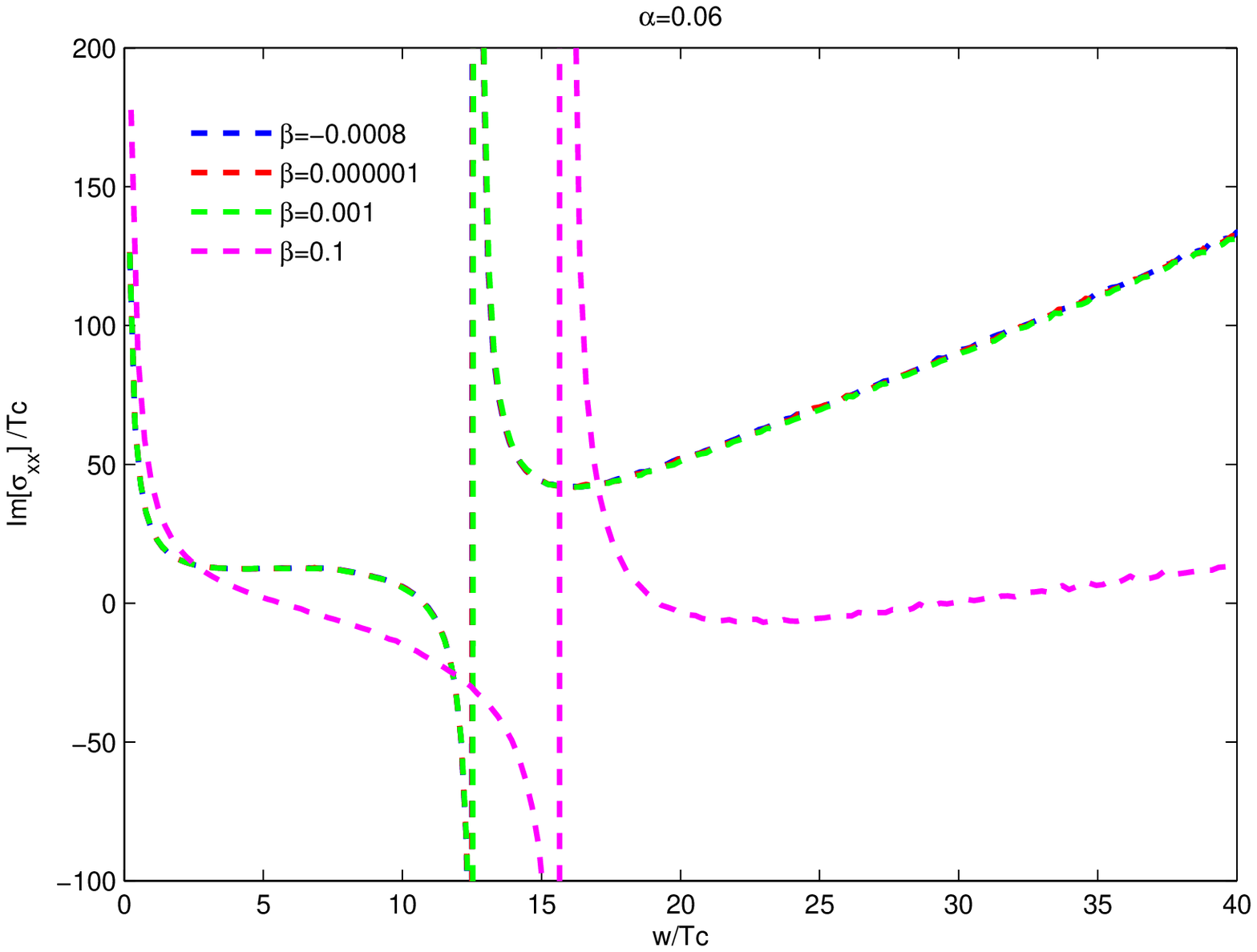}\hspace{0.0cm}
\caption{\label{fg}The conductivity $\sigma_{xx}$ for the p-wave
superconductors with a fixed $\alpha$. The left figure is for the
real part of $\sigma_{xx}$ while the right one is for the imaginary
part of $\sigma_{xx}$.}}
\end{figure}

\begin{figure}
\center{
\includegraphics[scale=0.4]{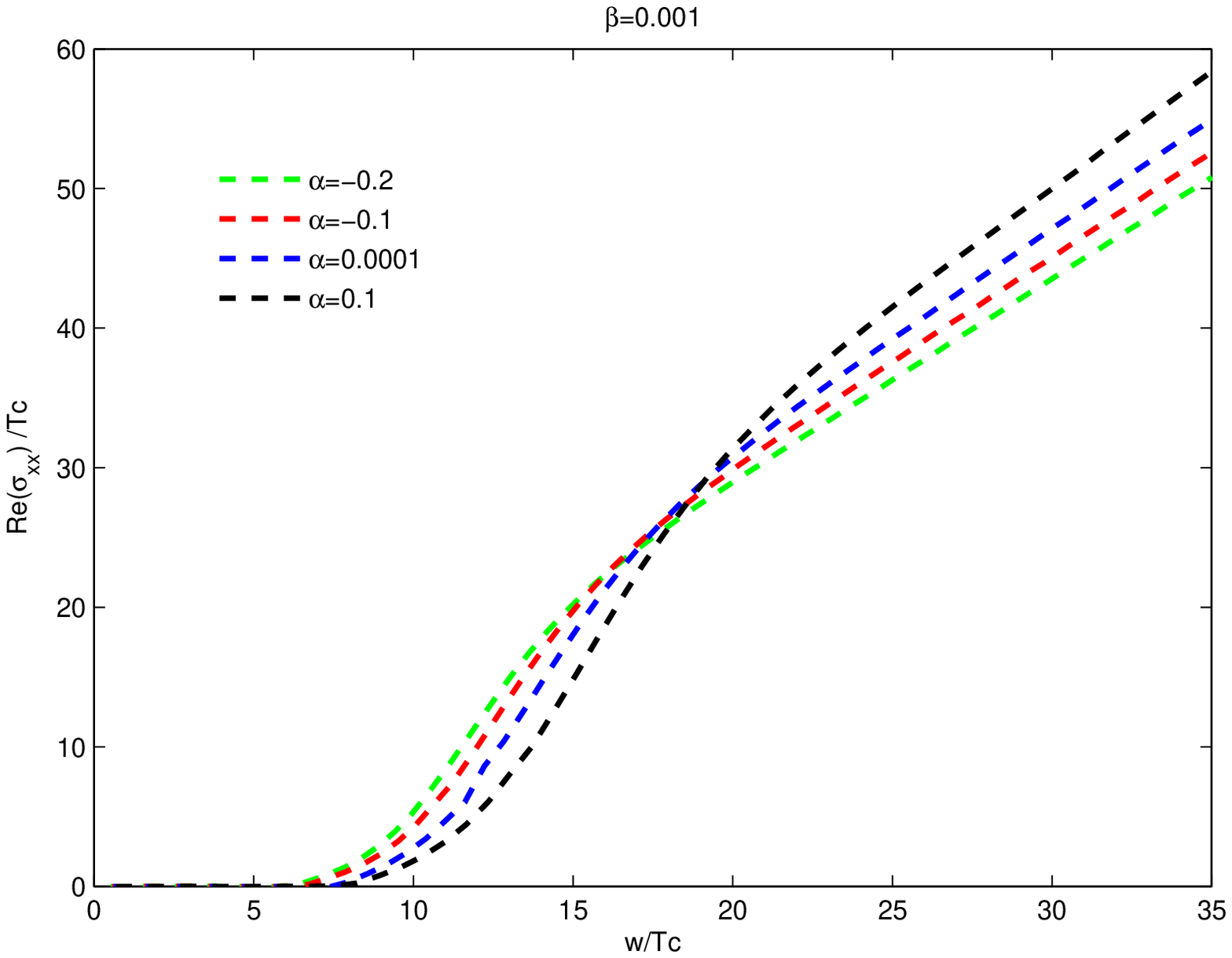}\hspace{0.0cm}
\includegraphics[scale=0.4]{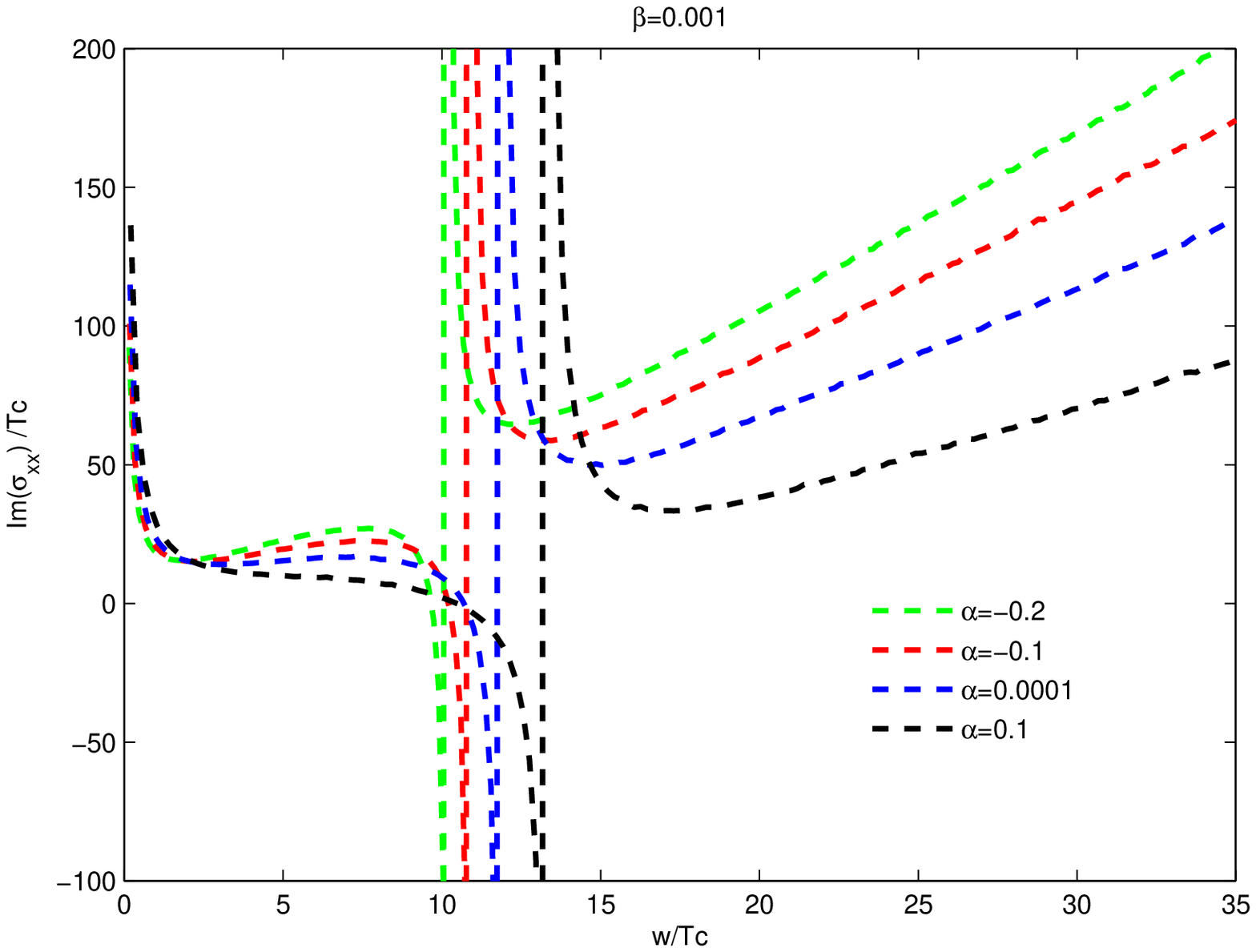}\hspace{0.0cm}
\caption{\label{fh}The conductivity $\sigma_{xx}$ for the p-wave
superconductors with a fixed curvature-cubed parameter
$\beta=0.001$. Similarly, the left figure is for the real part of
$\sigma_{xx}$ while the right one is for the imaginary part of
$\sigma_{xx}$.}}
\end{figure}

\begin{figure}
\center{
\includegraphics[scale=0.4]{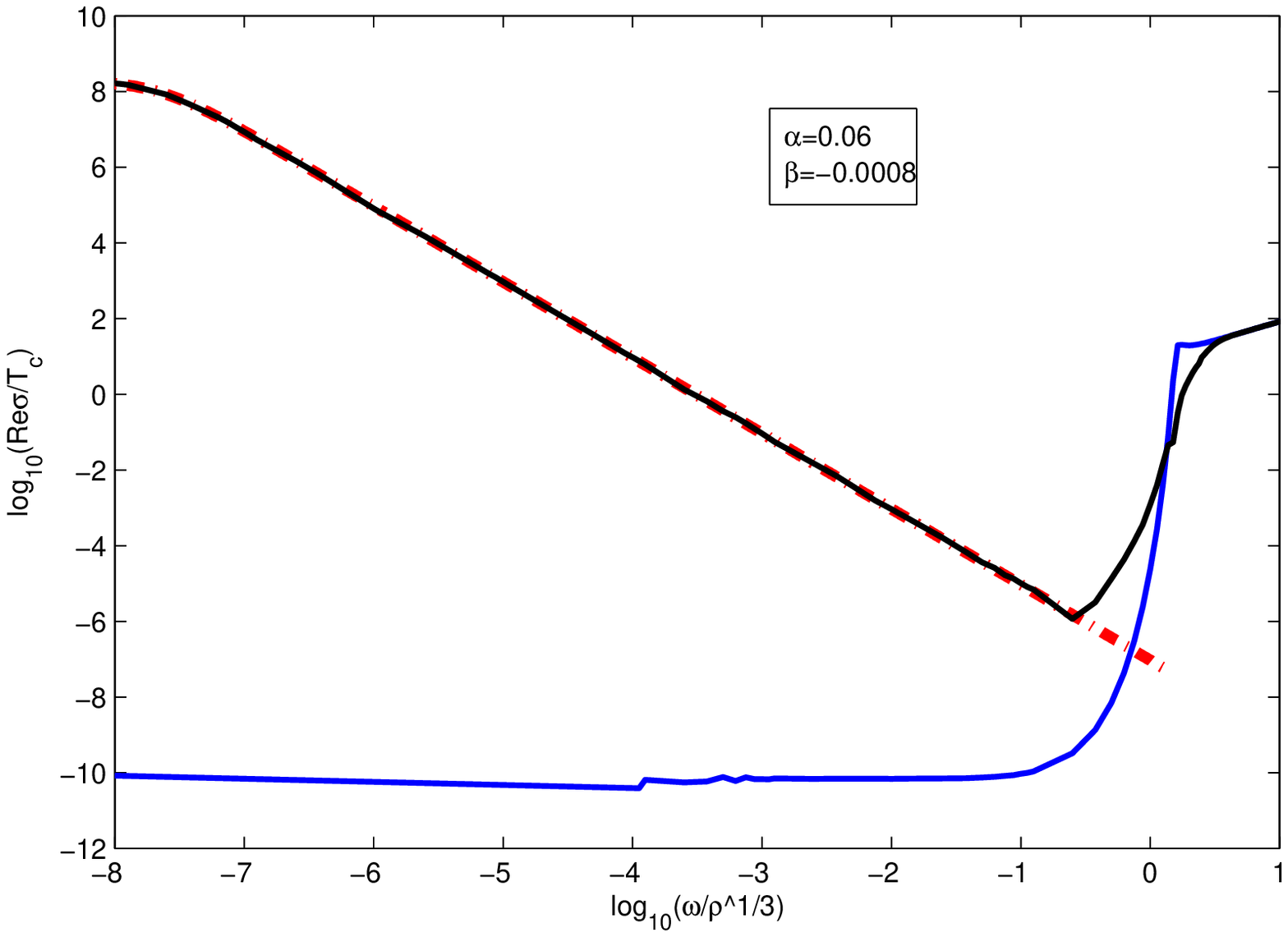}\hspace{0.0cm}
\includegraphics[scale=0.4]{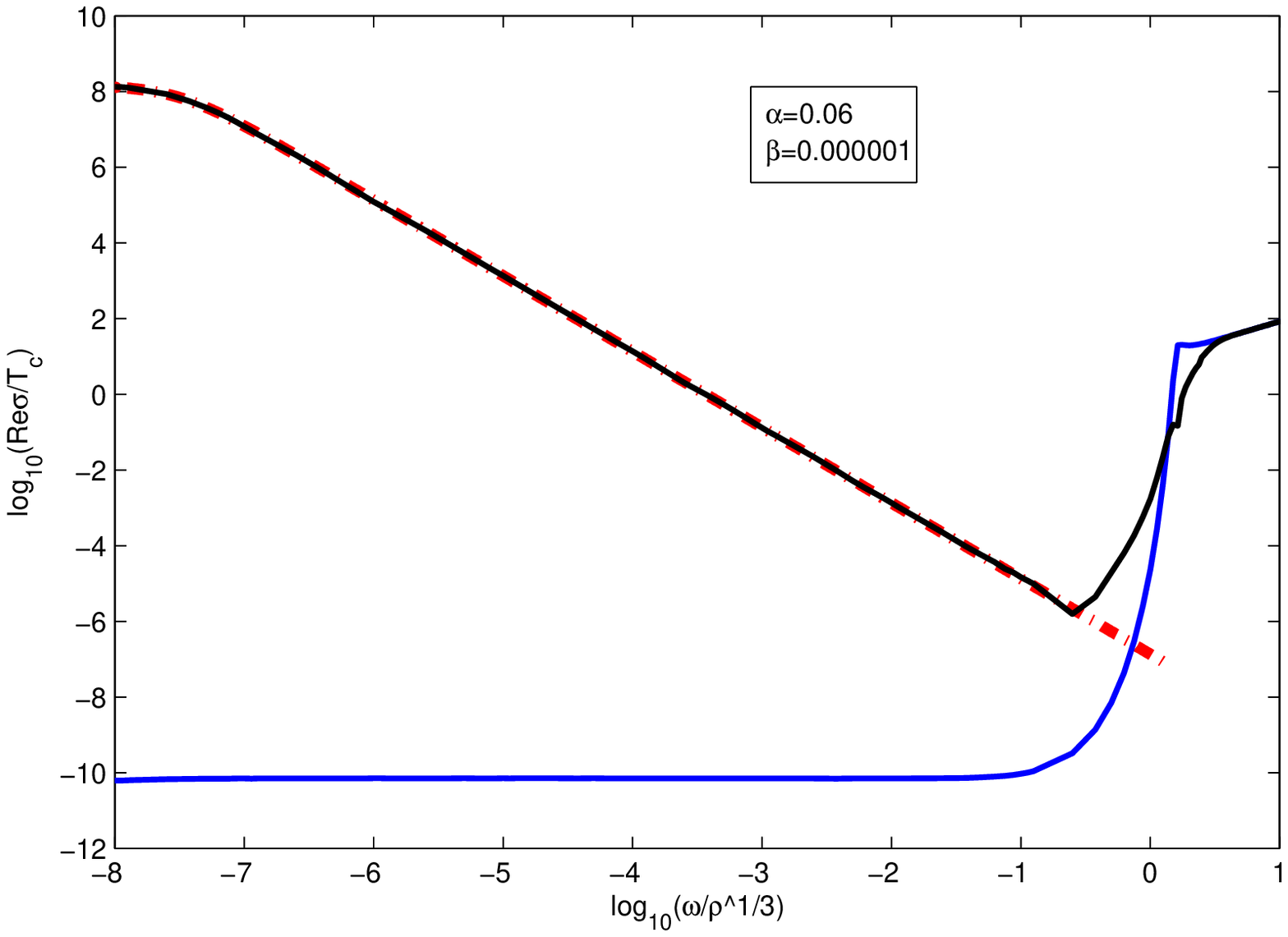}\\\hspace{0.0cm}
\includegraphics[scale=0.4]{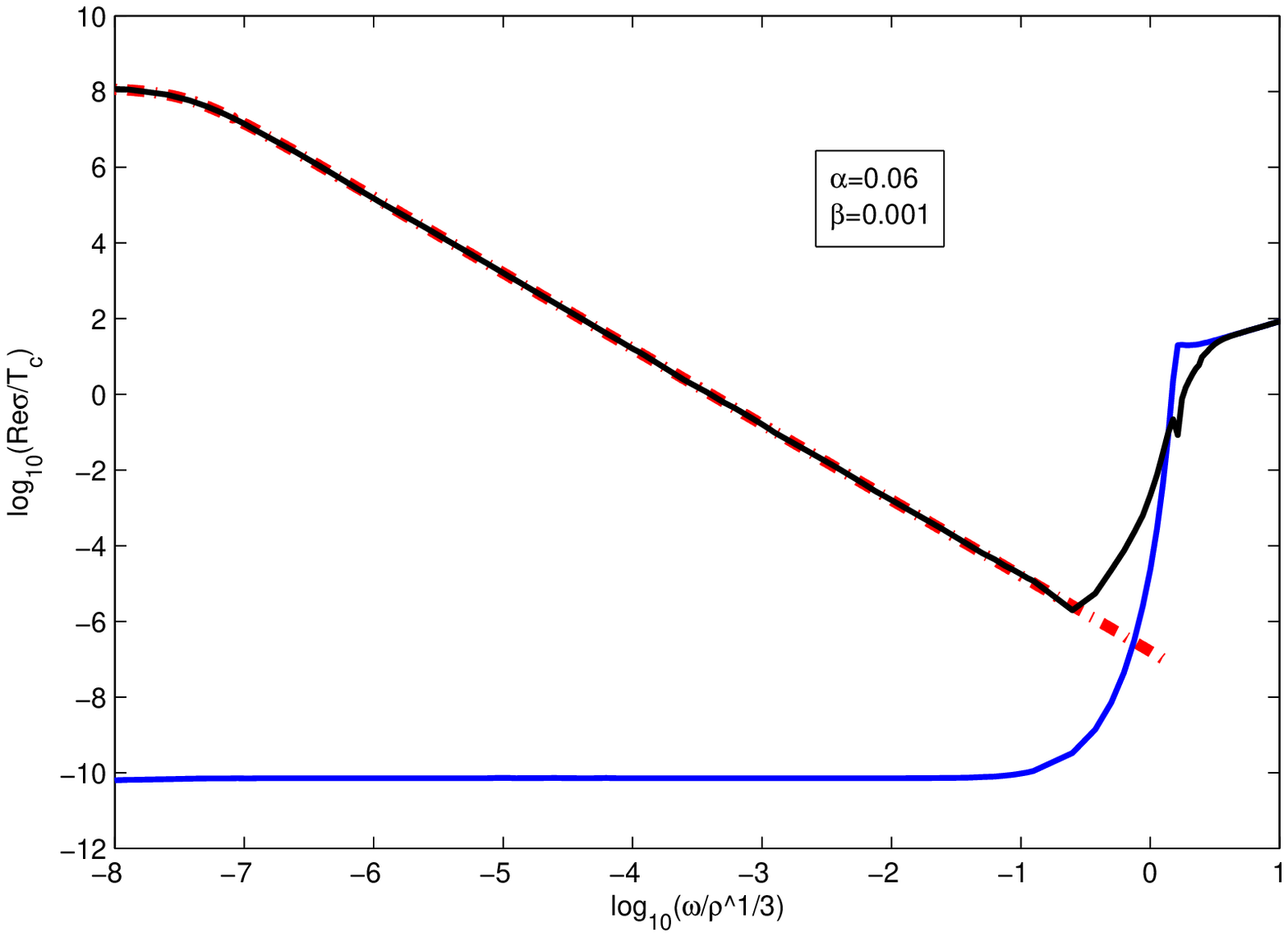}\hspace{0.0cm}
\includegraphics[scale=0.4]{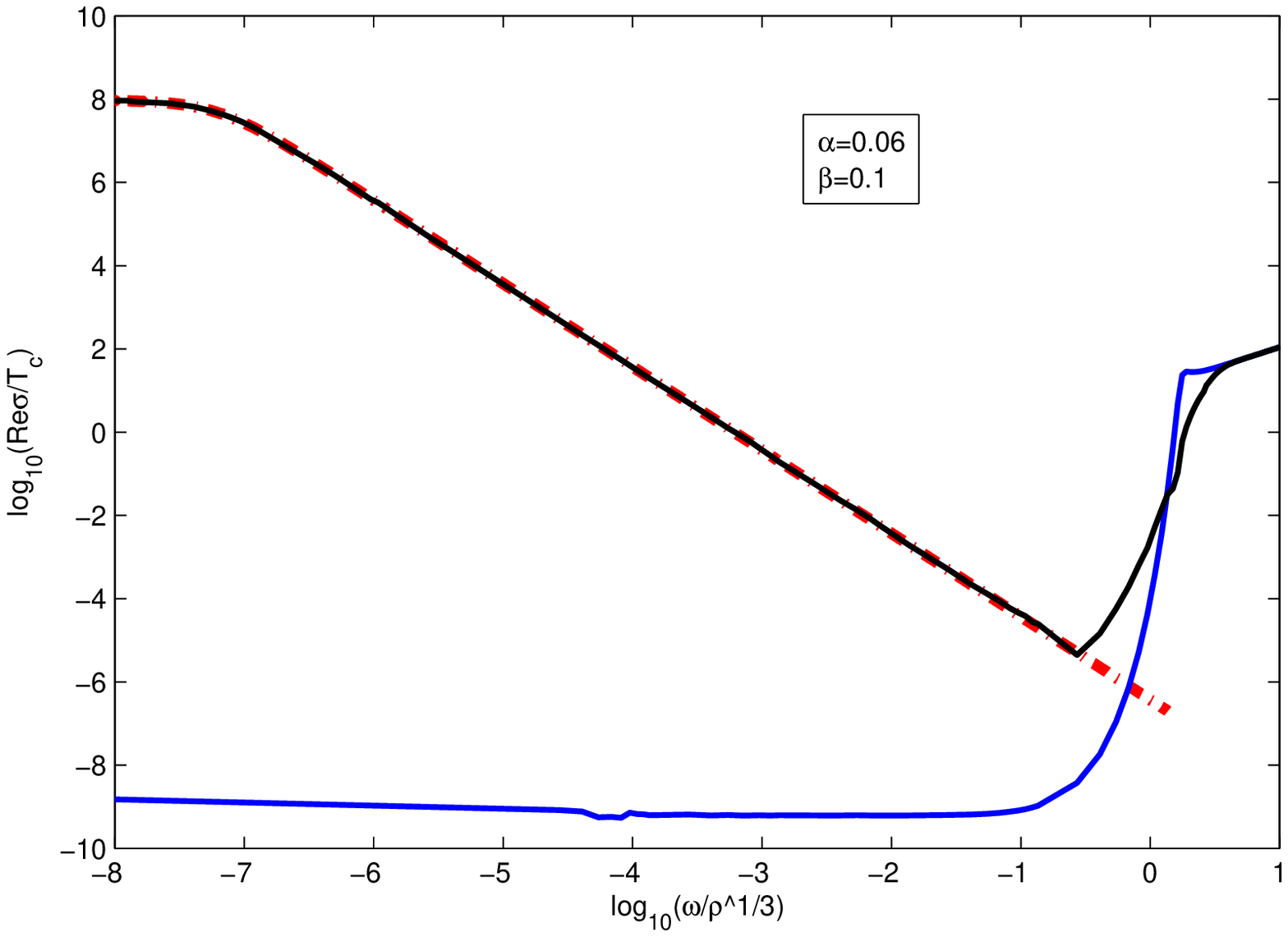}\hspace{0.0cm}
\caption{\label{fi}The relations between  $log_{10}[Re(\sigma/T_c)]$ and
$log_{10}[\omega/\rho^{(1/3)}]$ for fixed $\alpha=0.06$. The blue
and black lines are for $\sigma_{yy}$ and $\sigma_{xx}$
respectively, while the red ones are the fitting lines of
$\sigma_{xx}$ when $\omega$ is low enough.}}
\end{figure}
\begin{figure}
\center{
\includegraphics[scale=0.4]{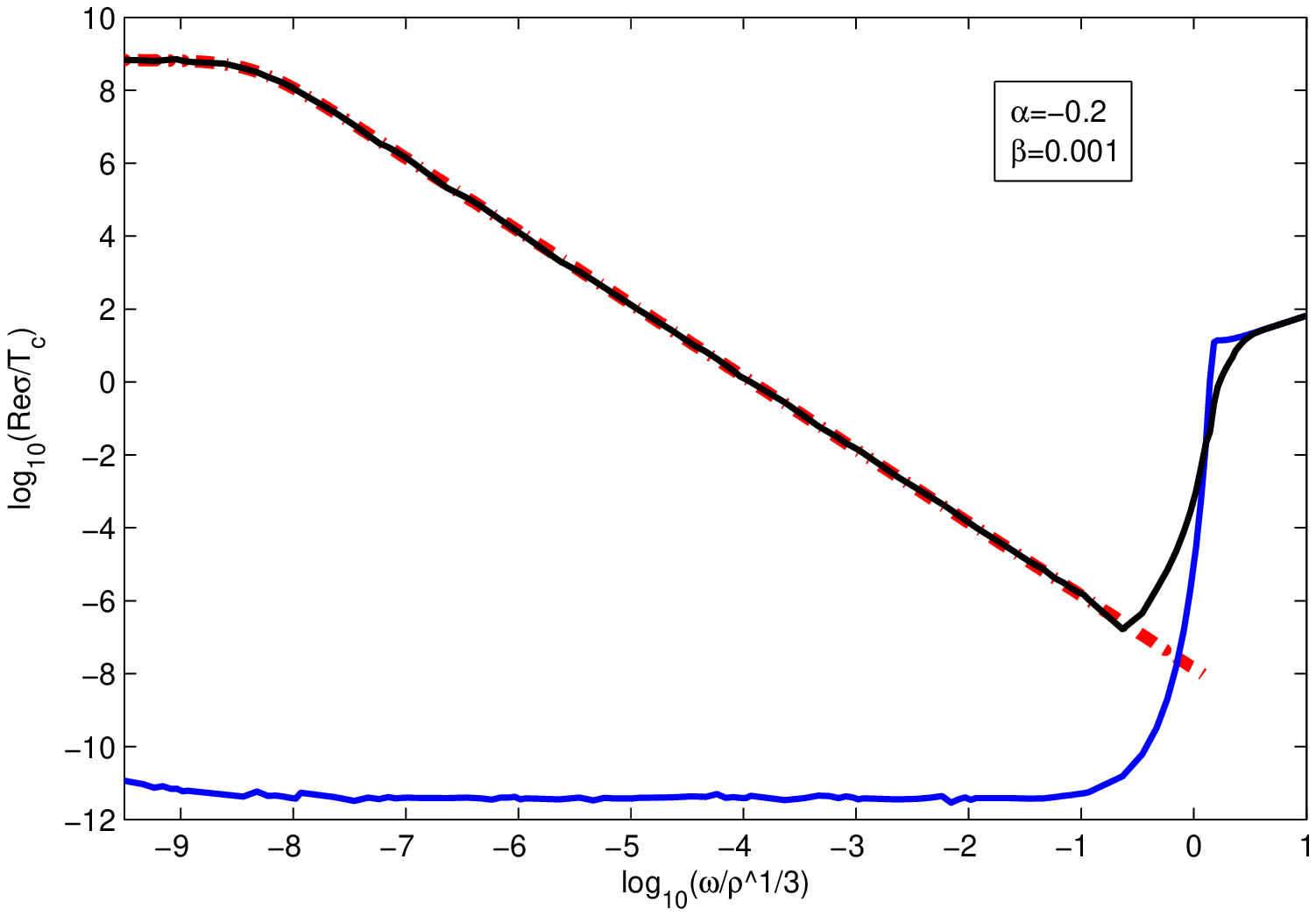}\hspace{0.0cm}
\includegraphics[scale=0.4]{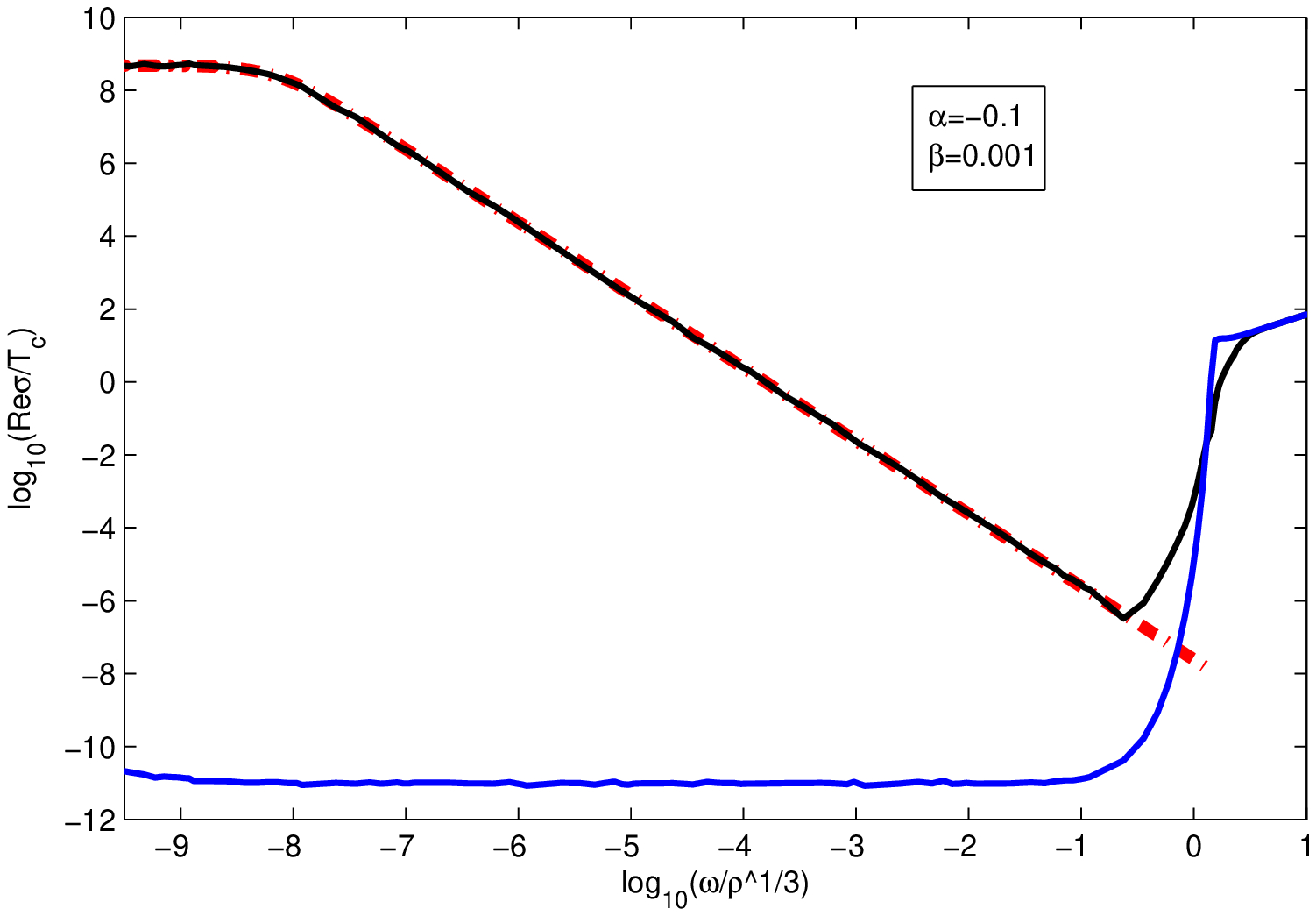}\\\hspace{0.0cm}
\includegraphics[scale=0.4]{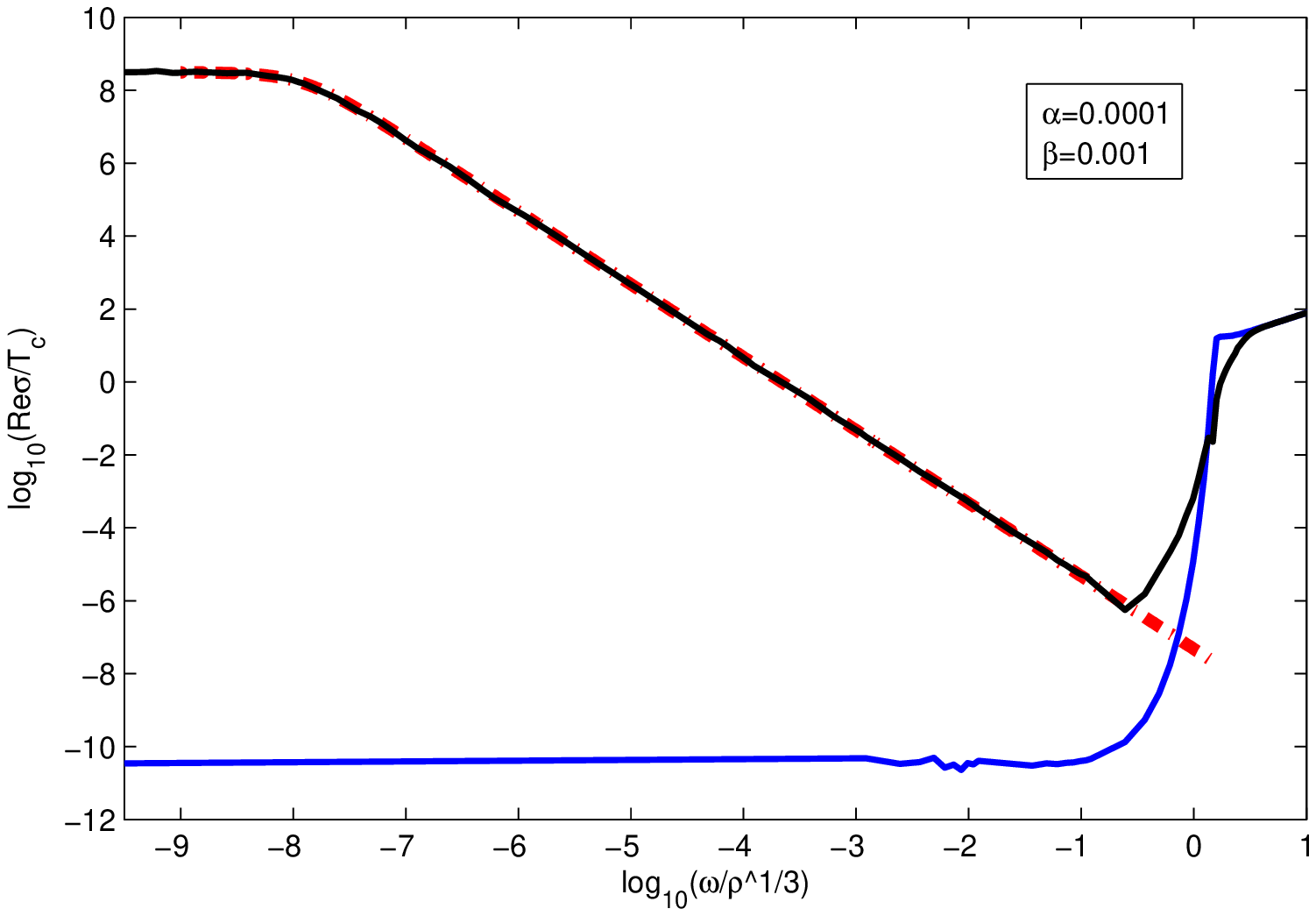}\hspace{0.0cm}
\includegraphics[scale=0.4]{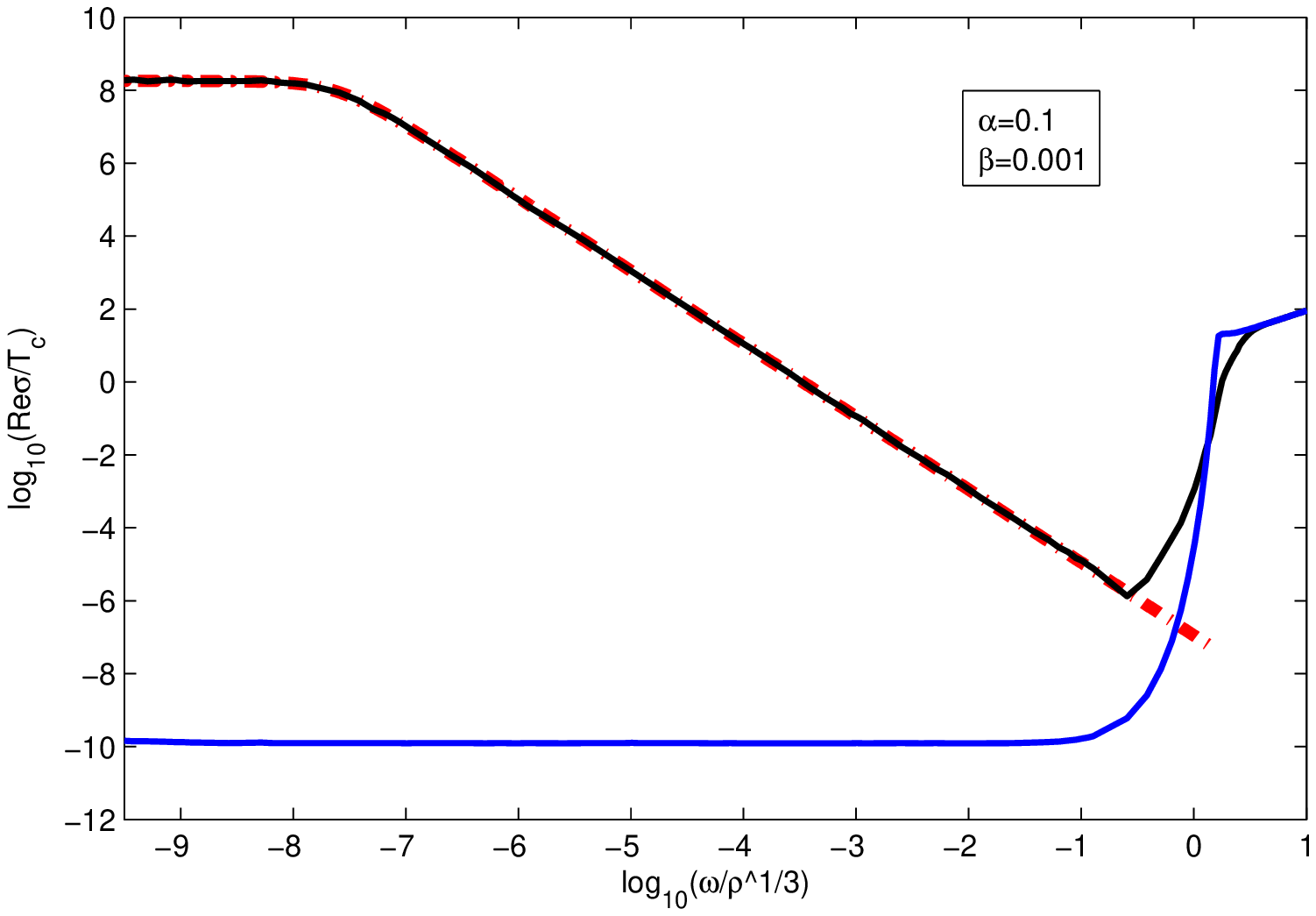}\hspace{0.0cm}
\caption{\label{fk}The relations between  $log_{10}[Re(\sigma/T_c)]$ and
$log_{10}[\omega/\rho^{(1/3)}]$ for fixed $\beta=0.001$. The blue
and black lines are for $\sigma_{yy}$ and $\sigma_{xx}$
respectively, while the red ones are the fitting lines of
$\sigma_{xx}$ when $\omega$ is low enough.}}
\end{figure}

\begin{table}

\begin{center}
\begin{tabular}{| c | c | c | c |}
\hline$\alpha$ &$\beta$ &$\sigma_0$/T &$\tau$T\\
\hline 0.06 & -0.0008 & $1.0213\times10^9$ & $1.8304\times10^6$\\
\hline 0.06 & 0.000001 & $7.5091\times10^8$ &$ 1.3064\times10^6$\\
\hline 0.06 & 0.001 &$6.4602\times10^8$ & $1.1072\times10^6$ \\
\hline 0.06 & 0.1 & $5.6827\times10^8$ & $6.38\times10^5$ \\
\hline -0.2 & 0.001& $3.735\times10^9$ &$8.804\times10^6$ \\
\hline -0.1 & 0.001 & $2.5534\times10^9$ & $5.588\times10^6$ \\
\hline 0.0001 & 0.001 & $1.5959\times10^9$ & $3.2288\times10^6$\\
\hline 0.1 & 0.001 & $8.7646\times10^8$ &$ 1.6036\times10^6$\\
\hline
\end{tabular}
\caption{The evaluation of $\sigma_0$ and $\tau$ and their
dependence on the values of coupling parameters.}
\end{center}
\end{table}
In order to understand the behavior of $\sigma_{xx}$, we
simultaneously solve the equation groups (\ref{j}), (\ref{k}),
(\ref{x}) as well as (\ref{y}). The numerical results of
$\sigma_{xx}$ are shown in FIG.\ref{fg} and FIG.\ref{fh}. In these two figures, we
find that the point of phase transition
 will also change when we choose different couplings, but the
behavior of the component $\sigma_{xx}$ is very different from that
of component $\sigma_{yy}$. Comparing with $\sigma_{yy}$, the real
part of $\sigma_{xx}$ ascends much more slowly. In the high
frequency limit, the imaginary part has a suppression effect with
the increase of the coupling parameters. On the other hand, in the
low frequency limit, $\sigma_{xx}$ behaves quite like that in Drude
model, and this analogy was firstly observed by Gubser in
\cite{Gubser}. Drude model is conventionally used to describe the
classical electron system, in which the real part of the
conductivity is featured by the relation

\begin{eqnarray}\label{bbb}
Re(\sigma)=\frac{\sigma_0}{1+\omega^{2}\tau^{2}},
\end{eqnarray}
where $\sigma_{0}=ne^2\tau/m$ is the DC conductivity, $n$, $e$,
$m$, $\tau$ are the electron density, charge, mass and the
relaxation time respectively.  Based on the Drude relation, we can
fit our numerical results to evaluate $\sigma_0$ and $\tau$ in
FIG. \ref{fi} and FIG .\ref{fk}. All the fittings are done under
the condition that the condensation is stable, namely, the value
of $T/\rho^{1/3}$ is small enough. We show the fitting results of
$\sigma_0$ and $\tau$ and their dependence on couplings parameters
$\alpha$ and $\beta$ in TABLE II. From the table, we find that
both of $\sigma_0$ and $\tau$  depend on the coupling parameters.

\section{discussions and conclusions}
In this paper we have constructed a p-wave holographic
superconductor model in quasi-topological gravity in the probe
limit. Firstly, we find that the superconducting condensation
becomes harder with the increase of the GB coupling and
curvature-cubed coupling. This is very similar to the case of the
s-wave model. Secondly, both anisotropic conductivities
$\sigma_{xx}$ and $\sigma_{yy}$ are studied. The numerical results
indicate that the behavior of $\sigma_{yy}$ at the low temperature
is quite similar to that in the s-wave model. More precisely, with
the increase of coupling parameters $\alpha$ or $\beta$ the ratio
$\omega_g/T_c\approx8$ becomes unstable and increases as well.
However, the conductivity $\sigma_{xx}$ behaves differently. Its
real part grows more slowly with the frequency, while its
imaginary part contains a spike. For both components, the
imaginary part is always suppressed by the increasing of couplings
in the large frequency limit. While in the low frequency limit,
our data of $\sigma_{xx}$ fits the Drude model very well, but the
values of DC conductivity as well as the relaxation time depend on
the coupling parameters.

It is worth pointing that due to the presence of higher curvature
corrections in quasi-topological gravity, besides the ghost modes
and naked singularity one should also be cautious of other
potential instabilities, such as the dynamical instability as
discussed in \cite{TS} for Lovelock gravity and the plasma
instability as discussed in \cite{Myers2,CEP}. Those potential
instabilities may further restrict the valid values of the
coupling parameters\cite{TS,Myers2,CEP}. We leave this open issue
for further investigation in future.

In the end of this paper we remark that it should be very worthy
to investigate p-wave superconductors when the back reactions are
taken into account in our model.  It is expected that both the
condensation and the charge transport would be corrected by the
effects of back reactions\cite{horowitz3,Ammon,Siani,cai2,ge}.
Moreover, inspired by recent progress on the holographic
non-fermion liquid and strange metals
\cite{Lee,HL,Polchinski,Cubraovic,Sachdev2,Wu}, it might be
possible to explore fermion system with a finite charge density in
the framework of the quasi-topological gravity. In these systems,
the fermion part near the horizon has a loop contribution to the
total two point correlator of the current, however, such an
$O(N^0)$ contribution dominates the dissipation of the
transport\cite{HL2}. Since in quasi-topological gravity the bulk
of spacetime may exhibit a richer structure of geometry due to the
higher order couplings, we propose that the dual CFTs would show
different behavior at low energy limit. This is expected to be
done in future.

\begin{acknowledgments}
We are grateful to Jian-Pin Wu, Hai-Qing Zhang and Hongbao Zhang for
reply and useful discussions. X. M. Kuang and Y. Ling is partly
supported by NSFC(Nos.10663001,10875057), JiangXi SF(Nos. 0612036,
0612038), Fok Ying Tung Education Foundation(No. 111008), the key
project of Chinese Ministry of Education(No.208072) and Jiangxi
young scientists(JingGang Star) program. W. J. Li is partly
supported by NSFC (No. 10975016). We also acknowledge the support by
the Program for Innovative Research Team of Nanchang University.
\end{acknowledgments}

\end{document}